\documentclass[prl,aps,twocolumn,showpacs,showkeys,preprintnumbers,longbibliography,superscriptaddress,amsmath,amssymb,floatfix,footinbib,a4paper,10pt]{revtex4-1}

\usepackage{graphicx}
\usepackage[usenames,dvipsnames]{color}
\usepackage{xcolor}
\usepackage{epsfig}
\usepackage{wasysym}
\usepackage{natbib}
\usepackage{bm}
\usepackage{siunitx}

\newcommand{\br}{{\bf r}}

\newcommand{\bj}{{\bf j}}
\newcommand{\be}{{\bf e}}

\newcommand{\bv}{{\bf v}}

\newcommand{\bV}{{\bf V}}
\newcommand{\bu}{{\bf u}}

\newcommand{\bG}{{\bf G}}

\newcommand{\Pitens}{{\mathsf{\Pi}}}
\newcommand{\Itens}{{\mathsf{I}}}
\newcommand{\bOmega}{{\boldsymbol{\Omega}}}

\newcommand{\vol}{d\mathcal{V}}

\newcommand{\eq}[1]{Eq.~(\ref{#1})}
\newcommand{\eqs}[1]{Eqs.~(\ref{#1})}

\newcommand{\Fref}[1]{Figure~\ref{#1}}
\begin{document}

\title{Self-motility of an active particle induced by correlations in
  the surrounding solution 
  }

\author{Alvaro Dom\'\i nguez}
\email{\texttt{dominguez@us.es}}
\affiliation{
  F{\'i}sica Te{\'o}rica, Universidad de Sevilla, Apdo.~1065, 
41080 Sevilla, Spain}
\affiliation{Instituto Carlos I de F{\'i}sica Te{\'o}rica y Computacional,
  18071 Granada, Spain}

\author{M.N.~Popescu}
\email{\texttt{popescu@is.mpg.de}}
\affiliation{Max-Planck-Institut f\"ur Intelligente Systeme, Heisenbergstr.~3, 
70569 Stuttgart, Germany}

\author{C.M.~Rohwer}
\email{\texttt{crohwer@is.mpg.de}}
\affiliation{Max-Planck-Institut f\"ur Intelligente Systeme, Heisenbergstr.~3, 
70569 Stuttgart, Germany}
\affiliation{IV.~Institut f\"ur Theoretische Physik,
  Universit\"{a}t Stuttgart, Pfaffenwaldring 57, D-70569 Stuttgart, Germany}
\affiliation{Department of Mathematics \& Applied Mathematics,
  University of Cape Town, 7701 Rondebosch, Cape Town, South Africa}

\author{S.~Dietrich}
\email{\texttt{dietrich@is.mpg.de}}
\affiliation{Max-Planck-Institut f\"ur Intelligente Systeme, Heisenbergstr.~3, 
70569 Stuttgart, Germany}
\affiliation{IV.~Institut f\"ur Theoretische Physik, Universit\"{a}t Stuttgart,
Pfaffenwaldring 57, D-70569 Stuttgart, Germany}

\date{November 4th, 2020}

\begin{abstract}
  Current models of phoretic transport rely on molecular forces
  creating a ``diffuse'' particle--fluid interface.  We investigate
  theoretically an alternative mechanism, in which a diffuse interface
  emerges solely due to a non-vanishing correlation length of the
  surrounding solution.  This mechanism can drive self-motility of a
  chemically active particle. Numerical estimates indicate that the
  velocity can reach micrometers per second.  The predicted
  phenomenology includes a bilinear dependence of the velocity on the
  activity and a possible double velocity reversal upon varying the
  correlation length.
\end{abstract}



\maketitle

Out-of-equilibrium behavior has been a research field of sustained
interest,  relevant to understanding the emergence of complexity
\cite{deMa84,SSG07,BBGR17,Walk17}. The last decade witnessed a rapidly
growing engagement with self-phoretic, chemically active particles as a new
paradigm thereof, which exhibits a wealth of phenomena, such as
micro-phase separation \cite{Palacci2013}, self-assembly of
super-structures \cite{Palacci2018}, and self-organized patterns of
collective motion \cite{Bechinger2019,Fischer2020}.  Additionally,
phoresis has proven to be an effective, versatile transport mechanism at the 
microscale \cite{Bocquet2019,Stone2019,Kawano2018}, leading to a significant
resurgence of interest also from a technological perspective.

Following the landmark review by Anderson \cite{Anderson1989}, the
phenomenon of phoresis of particles suspended in a fluid is
characterized by the presence of a driving thermodynamic field and by
a transition region (also called ``diffuse interface'') around the
particle. The forces relevant for the phoretic displacement act within
this region: its nonvanishing width gives rise to a response of the
fluid medium even though the net sum of the forces and torques
vanishes. Thus the particle is transported by the driving field while
the system (particle plus fluid) is mechanically isolated, which sets
this effect fundamentally apart from that of transport by an external
force.  The thermodynamic driving field usually comprises an
externally imposed gradient in one of the state-parameters (such as
composition, temperature, etc.) of the fluid, or of an electric field
in the case of a charged particle in an electrolyte. The alternative
case, in which the particle itself generates the driving field --- a
scenario also envisioned by Anderson \cite{Anderson1989} --- has
recently received much attention under the generic name of
``self-phoresis''
\cite{Golestanian2005,Ruckner2007,Golestanian2007,Moran2017}. Experimentally,
self-propelled particles have been realized most often in the form of
so-called Janus particles. In a typical scenario, such particles are
immersed in a multi-component fluid solution and promote catalytic
reactions which vary across their surfaces, leading to
self-electrophoresis \cite{Paxton2004,Fournier-Bidoz2005} or
self-chemiophoresis
\cite{Golestanian2005,Palacci2013,BrPo14,Simmchen2016,singh2017non}. Alternatively,
the particles behave as heat sources, leading to self-thermophoresis
through a single-component fluid \cite{Kroy2016} or
self-chemiophoresis via demixing of a critical binary liquid mixture
\cite{Volpe2011}.

A common ingredient in the models, which have so far addressed
phoretic \cite{Anderson1989} and self-phoretic
\cite{Golestanian2007,Wurger2015,samin2015self,Kroy2016,Moran2017,Brown2017,Liverpool2017}
phenomena, is a direct physical interaction between the particle and
the molecules of the surrounding fluid \footnote{The paradigmatic
  setups one has considered involve either electrostatic forces or
  dispersion forces between the particle and the components of the
  fluid solution \cite{Moran2017}.}. The corresponding non-vanishing
interaction range provides the aforementioned diffuse interfacial
region, where the \emph{internal} forces and torques act and induce
motion.  Here, we propose an alternative transport mechanism which
does not require this direct, extended interaction: instead, the
interfacial region is diffuse because the fluid exhibits a
\emph{non-vanishing correlation length}.  Although the phenomenology
driven by this mechanism could also be called ``phoresis'' (in
accordance with the definition proposed by Anderson
\cite{Anderson1989}), we coin the notion of ``correlation--induced
transport'' in order to distinguish it from the usual one, which
necessarily invokes a direct, spatially extended interaction.

Here we focus on the example of chemiophoresis, for which physically
insightful analytical expressions can be derived, in order to
illustrate the correlation--induced transport. However, the conceptual
framework discussed below is expected to exhibit a very general
applicability for other types of phoresis --- albeit significantly
more technically involved. For clarity and simplicity, we consider a
rigid, spherical particle (radius $R$), immersed in a solution, i.e.,
a solvent plus a single solute species (called ``the chemical'' in the
following).  The particle is active, i.e., it is a source or sink of
the chemical.  The two main approximations are a spatially and
temporally constant temperature and mass density of the solution, and
a slow motion of the particle \cite{SM}. Consequently, the state
of the solution can be characterized \cite{deMa84} by the
instantaneous stationary profiles of, e.g., the mass density of the
chemical or, equivalently, of its number density $n(\br)$ valid for
small P{\'e}clet numbers, and of the velocity field $\bu(\br)$ of the
solution valid for small Reynolds and Mach numbers. Thus, $n(\br)$ is
determined by the constraint of mass conservation for the chemical:
\begin{equation}
  \label{eq:chemstat}
  \frac{\partial n}{\partial t} = - \nabla \cdot\bj_\mathrm{diff} = 0 ,
  \qquad
  \bj_\mathrm{diff} = \Gamma \mathbf{f}. 
\end{equation}
Here $\Gamma$ is the mobility of the chemical within the solution, while 
$\mathbf{f}$ is the body force density acting on the chemical and which drives 
its diffusion. Likewise, $\bu(\br)$ follows from the balance between this body 
force and the fluid stresses, as expressed by the Stokes equation
for incompressible flow,
\begin{equation}
  \label{eq:stokes}
  \nabla\cdot\Pitens + \mathbf{f} = 0 ,
  \qquad
  \nabla\cdot\bu=0 ,
\end{equation}
in terms of the stress tensor
$\Pitens = \eta [ \nabla\bu + (\nabla\bu)^\dagger ] - \Itens \, p$
($\eta$ is the viscosity of the solution, $p$ the pressure field
enforcing incompressibility, and $\Itens$ the identity
tensor). Complementarily to the slow particle dynamics, one assumes
local equilibrium for the chemical \cite{Anderson1989}, so that
\begin{equation}
  \label{eq:f}
  \mathbf{f}(\br)= - n(\br) \nabla\mu(\br),
\end{equation}
in terms of its chemical potential $\mu$. The latter is modelled by
means of a free energy functional \cite{Wheeler1998} (we take
  spherical coordinates $(r,\theta,\varphi)$ with the origin at the
  center of the particle; $\vol$ denotes the volume element):
\begin{equation}
  \label{eq:H}
  \mathcal{H}[n] = 
  \int\limits_{|\br| > R}
    \vol
    \;\left[ 
    h(n) 
    + \frac{1}{2} \lambda^2 |\nabla n|^2 
    + n \, W(\br) 
  \right] \; ,
\end{equation}
so that
\begin{equation}
  \label{eq:mu}
  \mu(\br) = \frac{\delta\mathcal{H}[n]}{\delta n(\br)} 
  = h'(n(\br)) - \lambda^2 \nabla^2 n(\br) + W(\br) .
\end{equation}
Here, $h(n)$ is a local free energy density, and $W(\br)$ is a
potential energy (which has a non-zero and finite range) generated by
the particle \footnote{For an incompressible solution, $W$ is actually
  the difference in interaction of the particle with a molecule of
  solute and with a molecule of solvent, respectively
  \cite{deMa84}.}. In \eq{eq:H}, the term $\propto \lambda^2$ accounts
for the range of the molecular interactions between the dissolved
molecules of the chemical \footnote{More precisely, $\lambda$ can be
  directly related to the second moment of the attractive part of the
  pair potential between the molecules of the chemical.}. The relevant
control parameters of this model for the chemical are $\lambda$ and
the temperature (implicit in the definition of
$\mathcal{H}$). However, it turns out (see, c.f., \eq{eq:tildeV}) that
the results depend only on $\lambda$ and the correlation length
$\xi := \lambda/\sqrt{h''(n_0)}$ for a reference density $n_0$
\footnote{The free energy in Eq.~(\ref{eq:H}), with $W=0$ and
  linearized in $\delta n = n-n_0$, provides the equilibrium
  correlation function
  ${\protect{\langle}} \delta n(\br) \, \delta n(\mathbf{0})
  {\protect{\rangle}} \sim r^{-1} \mathrm{e}^{-r/\xi}$.}, without the
need to specify the dependence on temperature.

Finally, one has to specify boundary conditions: at infinity, a
homogeneous equilibrium state is recovered, i.e., a vanishing velocity
of the fluid solution and a constant chemical potential due to a
reservoir (thereby also fixing the density at a value $n_0$ such that
$h'(n_0)=\mu_0$):
\begin{equation}
  \label{eq:bcinfty}
  \bu(\br) \to 0 ,
  \;
  \mu(\br) \to \mu_0 ,
  \qquad
  \mathrm{as}
  \; |\br|\to \infty .
\end{equation}
On the surface of the particle, we impose a no-slip boundary condition
in terms of the translational ($\bV$) and angular ($\bOmega$)
velocities of the particle,
\begin{equation}
  \label{eq:noslip}
  \bu(\br) = \bV + \bOmega\times\br, 
  \qquad
  \mathrm{at}\; |\br|=R ,
\end{equation}
the boundary condition for the density, which follows from the surface
term of the variation of the free energy functional \footnote{In the
  limit $\lambda\to 0$, this boundary condition becomes actually
  irrelevant.},
\begin{equation}
  \label{eq:bcR}
  \lambda^2 \be_r\cdot\nabla n(\br)=0 ,
  \qquad
  \mathrm{at}\; |\br|=R , 
\end{equation}
and a prescribed current of the chemical, modelling the activity of the
particle as a source of the chemical:
\begin{equation}
  \label{eq:activity}
  \mathbf{e}_r \cdot \mathbf{j}_\mathrm{diff} (\br) = 
  q \mathbb{A}(\theta,\varphi) ,
  \qquad
  \mathrm{at}\; |\br|=R.
\end{equation}
The positive constant $q$ represents the production rate of the
chemical per area of the particle, and $\mathbb{A}(\theta,\varphi)$ is
the dimensionless activity, which can vary along the surface of the
particle.

Equations~(\ref{eq:chemstat}--\ref{eq:activity}), together with the
condition that the particle does not experience any external force or
torque, constitute a complete description for obtaining the velocities
$\bV$ and $\bOmega$ of the particle for a prescribed activity function
$\mathbb{A}$. If $\mathbb{A}=0$, the equilibrium state
$\mu(\br)=\mu_0$ is a solution of the equations, which implies $\bV=0$
and $\bOmega=0$. Thus, particle transport requires the non-equilibrium
imbalance introduced by the chemical activity. We address two
important aspects for solving this model. First, mechanical isolation
of the whole system ``particle + solution'' holds because, according
to \eq{eq:stokes}, the medium transmitting forces from the particle to
infinity is in local mechanical balance \cite{SM}.
Second, according to \eq{eq:stokes}, the incompressibility constraint
implies that the motion is driven by the spatially extended field
$\nabla\times\mathbf{f}$, which (because of mechanical isolation) is
the only source of flow vorticity \footnote{The force can be
  decomposed into potential (i.e., irrotational) and solenoidal (i.e.,
  divergence-free) components,
  $\mathbf{f} = \mathbf{f}_\mathrm{pot}+ \mathbf{f}_\mathrm{sol}$. The
  potential component,
  $\mathbf{f}_\mathrm{pot}=-\nabla \mathcal{P}_\mathrm{f}$, can be
  absorbed by the (auxiliary) hydrodynamic pressure field,
  $p \mapsto p + \mathcal{P}_\mathrm{f}$, that enforces
  incompressibility; see also Eqs.~(\ref{eq:Omegacurlf},
  \ref{eq:Vcurlf}) in Ref.~\cite{SM}.}.  Combining \eqs{eq:f} and
(\ref{eq:mu}), one obtains
\begin{eqnarray}
  \label{eq:curlf}
  \nabla\times\mathbf{f}
  &=
  & \nabla\mu \times\nabla n
    \nonumber\\
  &=
  & \nabla W \times \nabla n
    - \lambda^2 \nabla\times\left\{ \nabla\cdot \left[ (\nabla n)
    (\nabla n) \right] \right\}.~~
  \end{eqnarray}
Therefore, the transport, i.e., $\nabla\times\mathbf{f} \neq 0$, requires 
either an explicit interaction between the particle and the chemical, i.e., 
$W\neq 0$, or a deviation of the solution from ideality, i.e., $\lambda \neq 
0$, which is associated with a non-vanishing correlation length in the 
solution. To the best of our knowledge, all previous studies of phoresis have 
analyzed and identified the case $W\neq 0$ as the primary origin of particle 
transport. Here we address the opposite case ($W=0$, $\lambda\neq 0$), for 
which the model [Eqs.~(\ref{eq:chemstat}--\ref{eq:activity})] implies 
that the spatially homogeneous equilibrium state ($n(\br)=n_0$, $\bu(\br)=0$) 
would not be perturbed by the presence of a passive particle, i.e., $q=0$. This 
case isolates the effect of correlations and already provides distinctive 
observable predictions.

A complete analytical solution for the correlation--induced transport
velocity can be obtained if the deviations from homogeneity are small,
i.e., for small values (weak activity) of the Damköhler number
$\mathrm{Da} = q R \beta/(n_0 \Gamma)$, where $\beta$ is the inverse
thermal energy and $\Gamma/\beta$ is the diffusivity of the chemical:
the dimensionless ratio $\mathrm{Da}$ quantifies the activity--induced
``chemical crowding'' near the particle against the opposing effect of
diffusion. In this case, the spatial variation of $\eta$ and $\Gamma$
(via a possible dependence on the density $n$) can be
neglected. Consequently, the ``diffusion problem'' for the density
profile $n(\br)$, posed by Eqs.~(\ref{eq:chemstat}, \ref{eq:f},
\ref{eq:mu}, \ref{eq:bcinfty}, \ref{eq:bcR}, \ref{eq:activity}), is
decoupled from the velocity field and can be solved first on its
own. This can be carried out analytically, after linearizing in the
small quantity $n(\br)-n_0$.  Subsequently, one solves the
``hydrodynamical problem'', stated by Eqs.~(\ref{eq:stokes},
\ref{eq:bcinfty}, \ref{eq:noslip}), with the body force in
Eq.~(\ref{eq:f}) determined by the profile $n(\br)$ obtained in the
first step. Conveniently, the Lorentz reciprocal theorem allows one to
sidestep the full solution of the ``hydrodynamical problem'' and to
obtain expressions for $\bV$ and $\bOmega$ directly in terms of
$n(\br)$. Details of this procedure are collected in the Supplemental
Material \cite{SM}. By expanding the activity in terms of spherical
harmonics,
\begin{equation}
  \label{eq:alm}
  \mathbb{A}(\theta,\varphi) = \sum_{\ell=0}^{\infty}
  \sum_{m=-\ell}^\ell a_{\ell m} Y_{\ell m}(\theta, \varphi) ,
\end{equation}   
one obtains a compact expression for the correlation--induced
transport velocity:
\begin{align}
  \label{eq:tildeV}
  \bV =  
  & V_0 \,
    \sum_{\ell m} \sum_{\ell' m'} 
    a_{\ell m} a_{\ell' m'} \, 
    \nonumber
  \\
  &\times \left[ 
    g^{\perp}_{\ell \ell'} (\xi/R)
    \bG^{\perp}_{\ell m; \ell' m'} 
    + 
    g^{\parallel}_{\ell \ell'} (\xi/R)
    \bG^{\parallel}_{\ell m; \ell' m'}
    \right]
\end{align}
with the velocity scale
$V_0 := \mathrm{Da}^2 R^3/(6\pi \eta \beta^2 \lambda^2)$. In the
Supplemental Material we provide \footnote{See
  Eqs.~(\ref{eq:Gr}--\ref{eq:gp}) in Ref.~\cite{SM}.} the rather
lengthy expressions for the purely geometrical factors $\bG$, which
depend only on the mathematical properties of the spherical harmonics,
and for the dimensionless coefficients $g$, which fully incorporate
the dependence on the correlation length $\xi$ brought about by the
physical model. The superscripts $\perp$ and $\parallel$ denote the
contributions by the radial and tangential components of the body
force $\mathbf{f}$, respectively.  The angular velocity $\bOmega$
turns out to vanish identically (see below).

Several general conclusions can be extracted already from this
expression. Since \eq{eq:tildeV} is bilinear in the activity, the
velocity is invariant under the transformation
$a_{\ell m}\to -a_{\ell m}, \forall \ell,m$. This means that the
direction of translation is insensitive to whether the particle acts
as a source or a sink of the exchanged chemical. As shown by
\eq{eq:curlf}, a nonvanishing velocity requires a deviation from
equilibrium ($\nabla\mu\neq 0$), as well as a deviation from
homogeneity ($\nabla n\neq 0$). In the model of correlation--induced
transport, both requirements are enforced by the activity, which
explains the bilinear dependence. This is at variance with the case of
standard phoresis, which gives rise to a velocity linear in the
activity \cite{SM} because $W \neq 0$ usually suffices to bring about
a relevant inhomogeneity. Another consequence of the symmetric form of
the term $\propto \lambda^2$ in \eq{eq:curlf} is the absence of
chirality, so that $\bOmega=0$ \footnote{See the discussion of
  \eq{eq:tildeomega2} in Ref.~\cite{SM}.}. In the standard model, on
the contrary, the relative orientation of $\nabla n$ and $\nabla W$
will provide, in general, a preferred direction and sense of rotation,
thus inducing a non-vanishing angular velocity.
  
The geometrical factors $\bG^{\perp}_{\ell m; \ell' m'}$ and
$\bG^{\parallel}_{\ell m; \ell' m'}$ can be expressed in terms of the
Wigner 3j symbols \cite{SM}, which induce ``selection rules'': these
factors will vanish unless $|\ell-\ell'| = 1$ and $|m+m'|\leq
1$. Therefore, the sum in \eq{eq:tildeV} explicitly couples only pairs
of ``neighboring multipoles''. This result captures a major difference
as compared to standard phoresis. For instance, an activity pattern
lacking a dipole in \eq{eq:alm} generically yields
$\bV\neq 0$, while the same pattern of activity, together with a
spherically symmetric potential $W(r)$, does not yield phoresis in the
standard model \footnote{See Ref.~\cite{Golestanian2007}. There is
  also the complementary case: a purely dipolar pattern (i.e., the
  simplest model of a Janus particle with no net production) renders
  $\bV=0$ \cite{SM}, but in the standard model it yields a
  non-vanishing phoresis even for a spherically symmetric potential.}.

Another interesting feature of the transport velocity is revealed
  by considering the limit $\xi\ll R$: the deviations from
homogeneity are then localized within a thin layer (of thickness
$\propto \xi$) at the surface of the particle. In this case, the
velocity $\bV$ can be expressed as an integral over the surface of the
particle \cite{SM}:
\begin{equation}
  \label{eq:Vsurface}
  \bV \approx 
  -\frac{1}{4\pi} \int_0^\pi d\theta\; \sin\theta \int_0^{2\pi} d\varphi\;
  \bv_\mathrm{slip} (\theta,\varphi).
\end{equation}
The integrand can be interpreted as a slip velocity for the
``hydrodynamical problem'' \cite{Anderson1989}, which is proportional
to the tangential component of the density gradient,
\begin{equation}
  \label{eq:vslip}
  \bv_\mathrm{slip} (\br) := \mathcal{L}(\br) \, \nabla_{\parallel}
  n(\br) ,\qquad \mathrm{at}~|\br| = R,
\end{equation}
with a local coefficient of phoretic mobility:
\begin{equation}
  \label{eq:L}
  \mathcal{L}(\theta,\varphi) :=
  \frac{\mathrm{Da} \, \xi^3}{\eta\beta R} \,
  \mathbb{A}(\theta,\varphi) .
\end{equation}
Due to the explicit dependence of $\mathcal{L}$ on the activity (via
$\mathrm{Da}$), the present mechanism of self-propulsion cannot be
interpreted as ``passive'' transport in an external (albeit
self-induced) gradient, which is another significant deviation from
standard phoresis \footnote{Correlation--induced phoresis is also
  predicted for a passive particle in an external density gradient,
  yielding a transport velocity which is at least \emph{cubic} in this
  gradient~\cite{wip}. This might explain why it has not received
  attention, being overshadowed by the mechanism based on an extended
  interaction.}. Finally, we note that $V\sim\xi^5$ if $\xi\ll R$
because the tangential gradient vanishes $\propto \xi^2$ \cite{SM}.
In the opposite limit $\xi\gg R$, the velocity given by \eq{eq:tildeV}
approaches a finite value \footnote{This conclusion is not trivial
  because the density field $n(\br)$ does require a finite $\xi$ in
  order to satisfy the boundary condition $n\to n_0$ 
  at infinity \cite{SM}.}, which depends only on $V_0$ and
$a_{\ell m}$.

\begin{figure}[t!]
\includegraphics[width=0.45\textwidth]{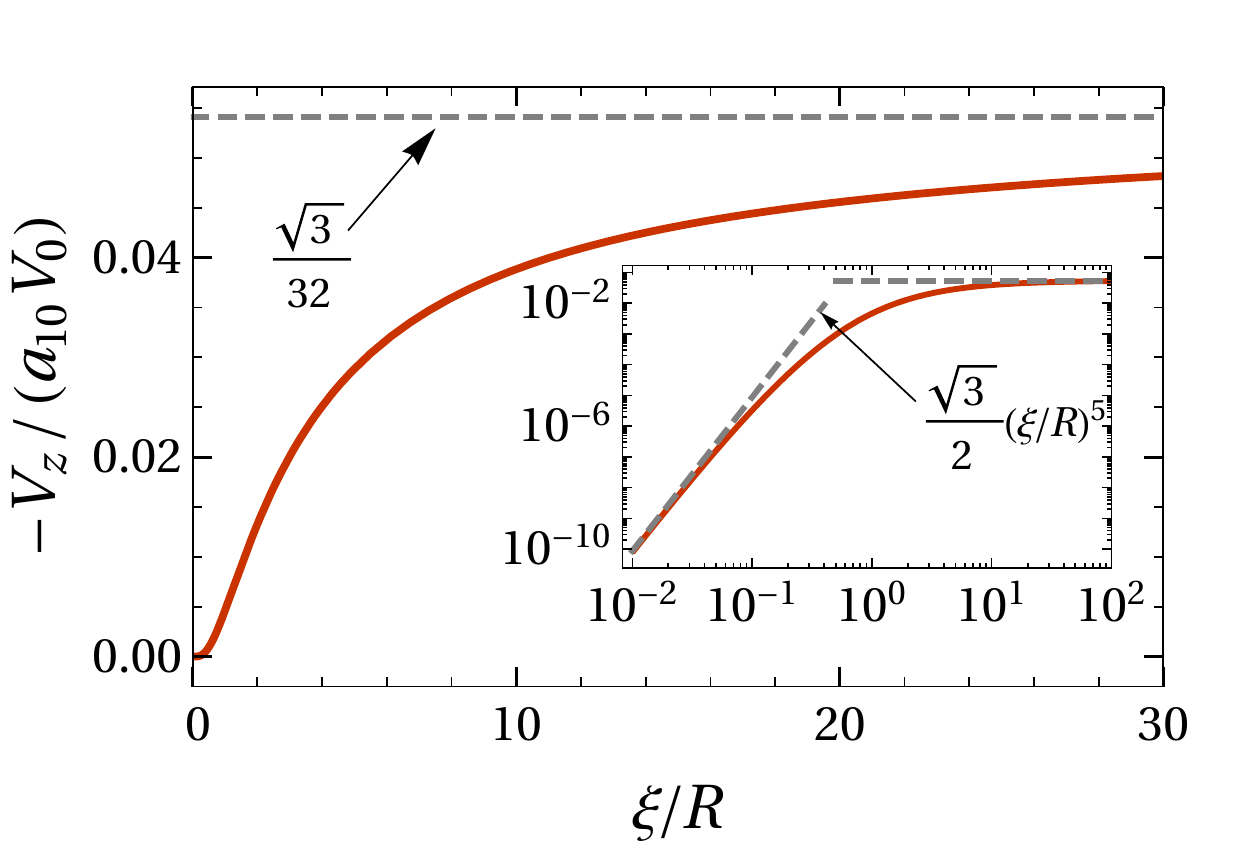}
\caption{Dimensionless $z$-component of the correlation--induced
  transport velocity as a function of the correlation length $\xi$ for
  the simplest case with polar symmetry (see the main text for
  details). The inset shows the same data on log-scales; the dashed
  lines indicate the asymptotic behaviors as
  $\xi \to 0~\mathrm{and}~\infty$, respectively.}
\label{fig:Vz}
\end{figure}

\begin{figure}[t!]
\includegraphics[width=0.48\textwidth]{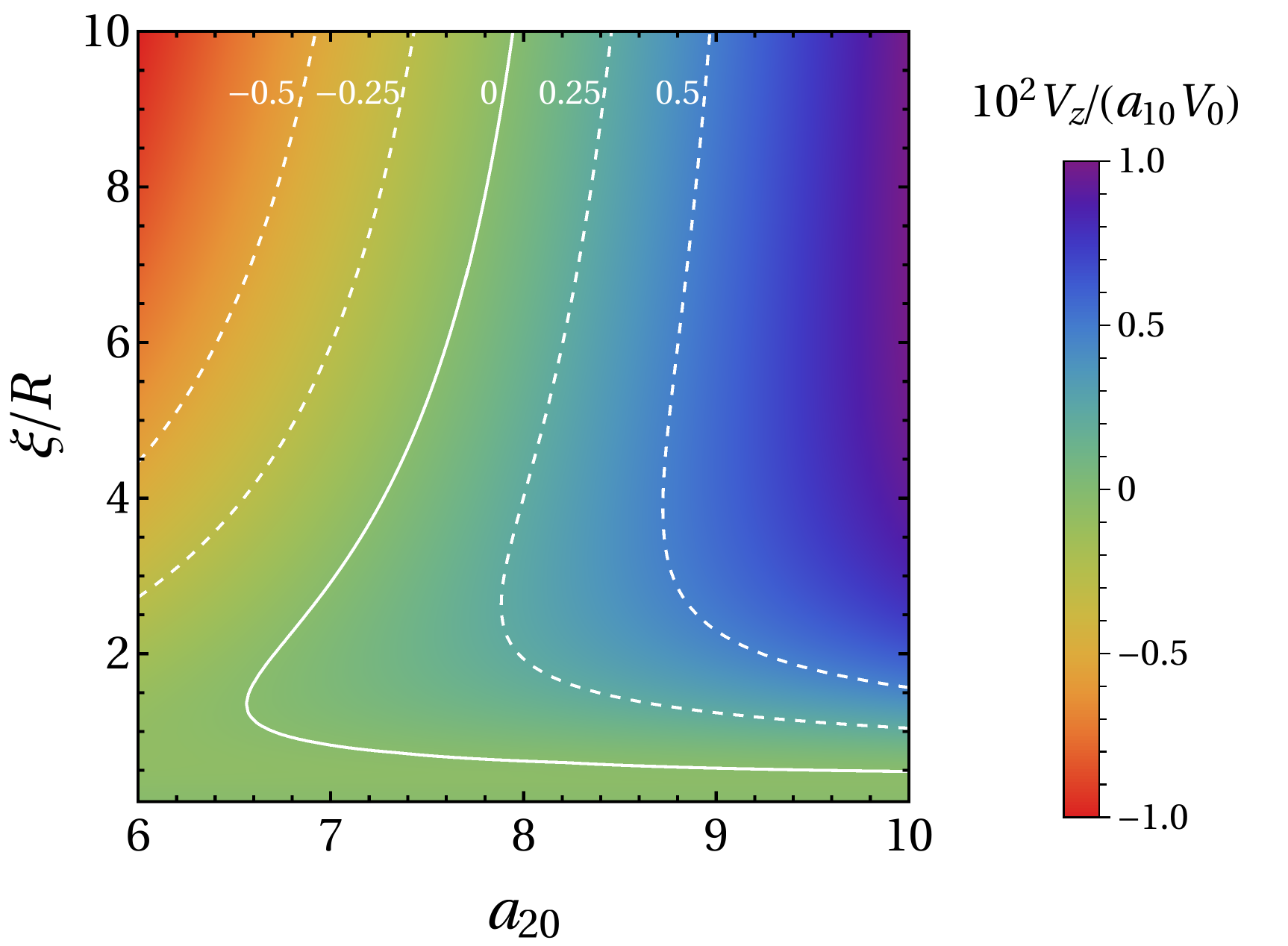}
\caption{Dimensionless $z$-component of the correlation--induced transport  
velocity as a function of the correlation length $\xi$ and of the quadrupolar 
activity parameter $a_{20}$ (see the main text for details). The white curves 
are contour lines.
}
\label{fig:reentr}
\end{figure}

As an illustration \cite{SM}, we study the simplest activity
distribution with polar symmetry by retaining only monopolar and
dipolar contributions: $a_{00} = 1$, $a_{1m} = a_{10} \,\delta_{m,0}$,
and $a_{\ell\geq2} =0$ in \eq{eq:alm}. For this case, \eq{eq:tildeV}
renders $\mathbf{V} = \be_z V_z$, $V_z = a_{10} V_0 f_1 (\xi/R)$, with
the function $f_1 (\xi/R)$ shown in \Fref{fig:Vz}.  The velocity
depends linearly on the dipole strength $a_{10}$ and vanishes in the
spherically symmetric limit $a_{10} \to 0$.  The asymptotic behaviors
for $\xi\ll R$ or $\xi\gg R$ are given by the general results
just discussed. If an additional quadrupolar activity contribution
$a_{20}\neq 0$ is included, \eq{eq:tildeV} gives
$V_z = a_{10} V_0 [ f_1 (\xi/R) + a_{20} f_2 (\xi/R) ]$, where $f_1$
and $f_2$ are monotonic functions of $\xi$ with opposite signs.
Strikingly, this implies that the sign of $V_z$ can be altered by
merely changing the correlation length $\xi$ (e.g., through
temperature), while keeping the particle properties (i.e., the
coefficients $a_{\ell m}$) fixed. This rich behavior is illustrated in
Fig.~\ref{fig:reentr}. For certain values of $a_{20}$, such as
$a_{20}=7$, one indeed observes a reentrant regime where $V_z$ is
negative at very small and very large $\xi$, but positive in an
intermediate range of correlation lengths.

These results allow one to estimate the magnitude of the velocity
under realistic conditions. From Fig.~\ref{fig:Vz}, upon taking
$a_{10}=1$, one obtains $V_z \approx V_0 (\xi/R)^5$ in the regime
$\xi/R < 1$. In order to estimate the velocity scale $V_0$ (see below
\eq{eq:tildeV}), we approximate the local free energy of the chemical
by the ideal gas expression $\beta h(n) = n (\ln n -1)$, so that
$\xi^2 = \beta n_0 \lambda^2$, and
$V_0 = (\mathrm{Da} \; R/\xi)^2 \, n_0 R/(6\pi\eta\beta)$. With the
typical values $R = 1~\mathrm{\mu m}$, $n_0 = 1~\mathrm{mM}$,
$\eta = 10^{-3}\;\mathrm{Pa \, s}$ (viscosity of water), and
$\mathrm{Da} = 10^{-1}$(as for, e.g., the Pt catalyzed decomposition
of hydrogen peroxide reported in Refs.~\cite{Paxton2004,Brown2017}),
one obtains $V_z \sim (10\,\xi/R)^3 \mathrm{\mu m/s}$ at room
temperature. For $\xi/R = 10^{-1}$, the predicted velocity would be
easily measurable; actually, the strong dependence on
$\xi/R$ 
provides a broad range of variation covering the values reported from
experiments for distinct types of systems
\cite{Ebbens2012PRE,BrPo14,Simmchen2016,singh2017non,das2020floor}.

In summary, we have studied a mechanism of phoresis driven by
correlations in the solution. Significant differences with respect to
the standard mechanism of phoresis become evident: (i) The
correlation--induced self-propulsion is bilinear in the activity, so
that for the same activity pattern the two mechanisms predict distinct
observable velocities. (ii) There is no self-rotation, regardless of
the activity pattern. (iii) The correlation--induced phoresis cannot
be understood as ``passive'' phoresis in an external (albeit
self-induced) driving field.  Already for simple activity patterns,
the self-propulsion velocity exhibits remarkable features, including
reentrant changes of sign obtained upon varying the correlation length
(see Fig.~\ref{fig:reentr}). Numerical estimates for realistic
conditions give values comparable to those observed experimentally,
which are usually interpreted within the framework of the standard
model of phoresis; correlation--induced phoresis provides an
additional, plausible mechanism for addressing these observations.
 
Thus, our formalism opens the door to studying the role which
correlations play in phoretic phenomena for a broad class of systems
and geometries. In the physically relevant situation in which
$\xi \neq 0$ and $W \neq 0$, both correlation--induced transport and
common phoresis would occur, including a coupling between correlations
within the fluid and the inhomogeneities induced by $W(\br)$. An
extension of our analytical model could address their relative
importance \cite{wip}. In this regard, the present study constitutes,
\textit{inter alia}, an important addition to the theoretical
machinery which is relevant for sorting and interpreting recent
results pertaining to self-propulsion due to demixing in a binary
liquid mixture (its order parameter playing the role of $n(\br)$).
For example, this encompasses the numerical analysis in
Ref.~\cite{samin2015self}, which incorporates nonlinear couplings (due
to a non-vanishing P{\'e}clet number) in order to explain the
emergence of self-motility, and the experimental observation of
velocity reversals for certain self-propelled particles
\cite{gomez2017tuning} following a change of illumination intensity
(which changes the temperature distribution in the fluid and,
correspondingly, the correlation length).

To conclude, we have identified and characterized a novel mechanism for 
self-phoresis of an active particle, which features potentially observable 
differences in comparison with the mechanism considered so far in the 
literature, and which can be controlled, e.g., via varying the temperature of 
the bath.

\begin{acknowledgments}
\label{Acknowledgments}
A.D.~acknowledges support by the Spanish Government through Grant
FIS2017-87117-P (partially financed by FEDER funds).
\end{acknowledgments}


%

\end{document}


\title{
  Self-motility of an active particle induced \\
  by correlations in the surrounding solution.  \\
  Supplemental Material\footnote{In this Supplemental
      Material, equations in the main text are referenced by those
    same numbers. Equations in the Supplemental Material are cited by
    a combination of the section number (in Roman numerals) and the
    equation number within the section (in Arabic numerals).}  }

\author{Alvaro Dom\'\i nguez}
\email{\texttt{dominguez@us.es}}
\affiliation{F\'\i sica Te\'orica, Universidad de Sevilla, Apdo.~1065, 
41080 Sevilla, Spain}
\affiliation{Instituto Carlos I de F\'\i sica Te\'orica y Computacional,
    18071 Granada, Spain}

\author{M.N.~Popescu}
\email{\texttt{popescu@is.mpg.de}}
\affiliation{Max-Planck-Institut f\"ur Intelligente Systeme, Heisenbergstr.~3, 
70569 Stuttgart, Germany}

\author{C.M.~Rohwer}
\email{\texttt{crohwer@is.mpg.de}}
\affiliation{Max-Planck-Institut f\"ur Intelligente Systeme, Heisenbergstr.~3, 
70569 Stuttgart, Germany}
\affiliation{IV.~Institut f\"ur Theoretische Physik, Universit\"{a}t Stuttgart,
Pfaffenwaldring 57, D-70569 Stuttgart, Germany}
\affiliation{Department of Mathematics \& Applied Mathematics,
  University of Cape Town, 7701 Rondebosch, Cape Town, South Africa}

\author{S.~Dietrich}
\email{\texttt{dietrich@is.mpg.de}}
\affiliation{Max-Planck-Institut f\"ur Intelligente Systeme, Heisenbergstr.~3, 
70569 Stuttgart, Germany}
\affiliation{IV.~Institut f\"ur Theoretische Physik, Universit\"{a}t Stuttgart,
Pfaffenwaldring 57, D-70569 Stuttgart, Germany}

\date{\today}

\maketitle

\tableofcontents

\newpage

 \section{General considerations concerning the balance of forces}
\label{sec:forceBalance}

The model system, which is analyzed in detail in the present study, consists of 
the particle and the fluid solution in which it is immersed; the latter 
comprises a solvent species and a solute species (denoted as ``chemical''). The 
chemical is involved in the activity of the particle (which changes the 
local composition of the solution). Here we discuss in more detail the 
conceptual framework of the ``force- and torque-free'' motility of the active 
particle, with an emphasis on the identification of the relevant forces. We 
first consider the force balance.

\begin{enumerate}
\item \textbf{The active particle:} The particle is assumed to drift
  at a constant velocity $\mathbf{V}$ and to rotate with a constant
  angular velocity $\bOmega$ relative to the fluid at infinity. We
  impose no-slip of the fluid solution at the impenetrable surface
  $\partial\mathcal{D}_\mathrm{part}$ of the particle (see
  Eq.~(\ref{eq:noslip}) in the main text). Since the particle is
    not acted upon by any external forces, the total force experienced
    by the particle consists only of the interaction with the fluid,
    \begin{equation}
    \label{eq:Fpart}
    \mathbf{F}^\mathrm{(part)} = 
    \oint\limits_{\partial\mathcal{D}_\mathrm{part}} \surf\;
    \be_n\cdot\Pitens(\br) 
     + \mathbf{F}^\mathrm{(part)}_\mathrm{chem}\,,
  \end{equation}
  where we take $\be_n$ to be the unit vector normal to
  $\partial \mathcal{D}_\mathrm{part}$ and pointing into the
  fluid. The first term involves the hydrodynamic stress tensor
  $\Pitens$ \cite{BrennerBook}, and the second term is the explicit
  force that the chemicals exert on the particle. In the overdamped
  regime, for which our model is formulated, the accelerations are
  negligible, so that the total force on the particle will vanish:
  \begin{equation}
    \label{eq:Fpartzero}
    \mathbf{F}^\mathrm{(part)} = 0 .
  \end{equation}
  
\item \textbf{The fluid solution:} In the regime of low Reynolds and
  Mach numbers, and assuming a steady state, the dynamics of the
  solution is described by the Stokes equations for incompressible
  creeping flow (see Eq.~(\ref{eq:stokes})). The integral version of
  this equation,
  \begin{equation}
    \label{eq:integralstokes}
    \int\limits_{\mathcal{D}} 
    \vol\;
    \nabla\cdot\Pitens(\br) 
    + \int\limits_{\mathcal{D}} 
    \vol\;
    \mathbf{f}(\mathbf{r})
    = 0,
  \end{equation}
  where $\mathcal{D}$ denotes the volume of the solution, describes
  the mechanical balance in bulk between the hydrodynamic forces and
  the force density field $\mathbf{f}(\mathbf{r})$, which represents
  the force acting on the molecules of the chemical in a volume
  element located at $\mathbf{r}$. The force $\mathbf{f} (\mathbf{r})$
  arises from interactions with other molecules of the chemical and
  from any external field created by the particle.
  
\item \textbf{Boundary condition at infinity and mechanical
    isolation:} At infinity we impose (see Eq.~(\ref{eq:bcinfty}) in
  the main text) that the fluid solution is at rest and that the
  perturbation of the density of the chemical also vanishes. This
  ensures that any local effect of the particle on its environment
  does not ``propagate too far''; actually, since the total system
  ``particle $+$ fluid solution'' is mechanically isolated, the
  combination of Eqs.~(\ref{eq:Fpart}--\ref{eq:integralstokes})
  renders
  \begin{eqnarray}
    \label{eq:Ftot}
    0 = \mathbf{F}^\mathrm{(total)} 
    & = 
    & \textrm{force on particle}
     + \textrm{force on fluid solution}  
    \nonumber\\
    & = 
    & \mathbf{F}^\mathrm{(part)} 
      + \int\limits_{\mathcal{D}} 
      \vol\;
      \nabla\cdot\Pitens(\br) 
      + \int\limits_{\mathcal{D}} 
      \vol\;
      \mathbf{f}(\mathbf{r})
      \nonumber\\
    & =
    & \mathbf{F}^\mathrm{(part)}_\mathrm{chem} 
      + \oint\limits_{\partial\mathcal{D}_\infty}
      \surf\; (-\be_n)\cdot\Pitens(\br) 
      + \int\limits_{\mathcal{D}} 
      \vol\;
      \mathbf{f}(\mathbf{r}) ,
  \end{eqnarray}
  where the volume $\mathcal{D}$ is bounded by the surface
  $\partial \mathcal{D}_\mathrm{part}$ of the particle and the surface
  $\partial \mathcal{D}_\infty$ at infinity, with the unit normal
  $\be_n$ oriented towards the fluid. Due to the principle of
  action--reaction, the ``chemical'' force experienced by the particle
  is balanced by the total force experienced by the chemical in the
  fluid solution:
  \begin{equation}
    \label{eq:fchembalance}
    0 = \mathbf{F}^\mathrm{(part)}_\mathrm{chem} 
    +\int\limits_{\mathcal{D}} 
    \vol\;
    \mathbf{f}(\mathbf{r}) .
  \end{equation}
  Thus one concludes from \eq{eq:Ftot} that
  \begin{equation}
    \label{eq:nohydroshear}
    \oint\limits_{\partial\mathcal{D}_\infty}
    \surf\; \be_n\cdot\Pitens(\br) = 0.
  \end{equation}
  Therefore, at infinity the flow must decay sufficiently fast so
  that the integrated hydrodynamic stress vanishes far from the
  particle. This would not be the case if the particle would
  experience an external force.
\end{enumerate}

The argument can be repeated, without conceptual changes, in order to
address the torque balance. In particular, it is found that the total
torque exerted on the particle can be written as the sum of
hydrodynamical and chemical contributions:
\begin{equation}
  \label{eq:torquepart}
  \boldsymbol{\tau}^\mathrm{(part)} =
  \int\limits_{\mathcal{D}} 
  \vol\;
  \br\times\left[\nabla\cdot\Pitens(\br) \right]
  + \int\limits_{\mathcal{D}} 
  \vol\;
  \br\times\mathbf{f}(\mathbf{r})
  = 0 ,
\end{equation}
which vanishes due to the balance condition expressed in
\eq{eq:integralstokes}. Additionally, the integrated hydrodynamic
torque at infinity vanishes:
\begin{equation}
  \label{eq:nohydrotorque}
  \oint\limits_{\partial\mathcal{D}_\infty}
  \surf\; \br\times \left[\be_n\cdot\Pitens(\br)\right] = 0.
\end{equation}
Recalling Eqs.~(\ref{eq:f}, \ref{eq:mu}), the force density can be
written as
\begin{equation}
     \label{eq:fexpand}
     \mathbf{f} = - n \nabla\mu 
     = -n \nabla W 
     - n h''(n) \nabla n
     + \lambda^2 n \nabla( \nabla^2 n)
     = - n \nabla W + \nabla\cdot \Pitens_\mathrm{chem}
\end{equation}
in terms of a chemical (or osmotic) stress tensor (sometimes called the 
Korteweg stress tensor) \cite{Wheeler1998}
 \begin{equation}
   \label{eq:Pichem}
     \Pitens_\mathrm{chem}
     := \lambda^2 \left[ n \nabla \nabla n 
       - \frac{1}{2} |\nabla n|^2 \, \Itens 
     \right] 
     - \Itens \; p_\mathrm{chem}(n) ,
   \end{equation}
   where the function
   \begin{equation}
     \label{eq:pchem}
     p_\mathrm{chem}(n) := n h'(n) - h(n) = \int\limits_0^n d\hat{n} \; 
       \hat{n}\, h''(\hat{n}) 
     \end{equation}
represents 
the osmotic pressure by the chemical in the limit $\lambda\to
0$. (Note that for $n \to 0$ (ideal gas) one has
  $h(n) \to k_B T n \ln n$ and $p_\mathrm{chem} \to n k_B T$.)
Accordingly, the force balance arguments above can be formulated in an
equivalent manner in terms of the chemical stress tensor. The total
force on the particle, as given by Eqs.~(\ref{eq:Fpart}) and
(\ref{eq:fchembalance}), turns into
   \begin{eqnarray}
     \mathbf{F}^\mathrm{(part)}
     & = 
     & \int\limits_{\mathcal{D}} 
     \vol\;
     n \, \nabla W
     + \oint\limits_{\partial\mathcal{D}_\mathrm{part}} \surf\;
     \be_n\cdot \left( \Pitens 
       + \Pitens_\mathrm{chem}
       \right)
         \nonumber
     \\
     & =
     & \int\limits_{\mathcal{D}} 
     \vol\;
       n \, \nabla W
       - \oint\limits_{\partial\mathcal{D}_\mathrm{part}} \surf\;
       \be_n \left[ p + p_\mathrm{chem}
       + \lambda^2\left( \frac{1}{3} n \nabla^2 n
       - \frac{1}{2} |\nabla n|^2
       \right)
       \right]
       \nonumber
     \\
     &
     & \mbox{}
       + \oint\limits_{\partial\mathcal{D}_\mathrm{part}} \surf\;
       \be_n\cdot \left\{
       \eta \left[ \nabla \bu + (\nabla \bu)^\dagger \right]
       + \lambda^2 \, n \left[ \nabla \nabla n 
        - \frac{1}{3} \, \Itens \, \nabla^2 n
        \right]
       \right\} ,
   \end{eqnarray}
   after using the explicit expressions in Eqs.~(\ref{eq:Pichem}) and
   (\ref{eq:pres_tens}) for the chemical and the hydrodynamical stress
   tensor, respectively.
We note that the total osmotic pressure, represented by the 
term $p_\mathrm{chem}$ and the trace of the correlation--induced contribution, 
is actually irrelevant for the dynamics, as it can be absorbed into the anyhow 
unspecified dynamic pressure $p$, which enforces the incompressibility of 
the fluid flow. Thus the only forces being effective in driving transport 
are the distant interaction represented by $W$, and the local, 
correlation--induced osmotic stress, which is proportional to $\lambda^2$. 
This is a restatement of the conclusions already extracted from \eq{eq:curlf} 
in the main text.

\section{The diffusion problem\label{sec:chem}}

The ``diffusion problem'' for the density profile $n(\br)$ is given by
Eqs.~(\ref{eq:chemstat}, \ref{eq:f}, \ref{eq:mu}, \ref{eq:bcinfty},
\ref{eq:bcR}, \ref{eq:activity}) with $W=0$. This set of equations was
derived under the assumption of slow particle motion so that, on the
time scale of the particle motion, the distribution of chemicals can
be considered to be in the stationary state associated with the
instantaneous particle position. More specifically, if the particle
moves with velocity $V$, changes in the configuration will be
noticeable after a time of the order $t_\mathrm{part} \sim R/V$ of
particle displacement by a sizeable fraction of itself. On the other
hand, the diffusive dynamics of the chemicals is characterized by a
diffusion coefficient $D_\mathrm{chem}=\Gamma/\beta$, in terms of the
chemical mobility $\Gamma$ and the inverse thermal energy $\beta$, so
that the time it takes the chemical to smooth out perturbations over a
length scale $R$ is of the order of
$t_\mathrm{chem} \sim R^2/D_\mathrm{chem}$. Consequently, the
dimensionless ratio
\begin{equation}
  \frac{t_\mathrm{chem}}{t_\mathrm{part}} \sim \frac{V R}{D_\mathrm{chem}}
\end{equation}
quantifies what it means to be ``slow'' from the perspective of the
dynamics of the chemical. With the typical numbers
$D_\mathrm{chem}\sim 10^{-9} \, \mathrm{m^2/s}$ for molecular
diffusion and the observed self-phoretic velocity
$V\sim \mu\mathrm{m}/s$ for micron--sized particles, this gives a
ratio $\approx 10^{-3}$. Thus, there is a clear separation of time
scales and the diffusion problem may be treated to be stationary. An
analogous reasoning allows one to quantify the relevance of the
convective transport of chemicals by the hydrodynamic flow, which
would be described by a term $\bu\cdot\nabla n$ in \eq{eq:chemstat}:
if the gradients in the chemical distribution are associated with a
length scale $L$, the diffusion is able to smooth out the distortions
induced by the hydrodynamic flow, characterized also by the velocity
scale $V$ of the particle motion, provided the time ratio
$V L/D_\mathrm{chem}$
(which is actually the P{\'e}clet number) is small. An estimate of the
length $L$ can be either the particle radius $R$ or an intrinsic
length related to the chemical distribution, e.g., the correlation
length $\xi$ (see, c.f., \eq{eq:xi}) or the spatial range of the
interaction potential $W(\br)$. For the numbers quoted above, the
P{\'e}clet number is small as long as the scale $L$ lies below the
millimeter ($\sim 10^3 R$).

The diffusion problem can be solved analytically in the limit that the
deviations $\delta n(\br) = n(\br)-n_0$ from homogeneity are small,
with $n_0$ being fixed according to $\mu_0= h'(n_0)$. As it will be
shown, these deviations can be quantified in terms of the
dimensionless ratio
\begin{equation}
  \label{eq:Da}
  \mathrm{Da} := \frac{q R \beta}{n_0 \Gamma}
  = \frac{q R}{n_0 D_\mathrm{chem}} ,
\end{equation}
which, in the language of chemical kinetics, is the Damk{\"o}hler
number of the chemical.  According to the boundary condition in
Eq.~(\ref{eq:activity}), which models the effect of activity, the
ratio $q/n_0$ can be interpreted as the effective velocity of influx
current of chemicals due to activity. On the other hand, as we have
seen, $t_\mathrm{chem}\sim R^2/D_\mathrm{chem}$ is the characteristic
time for diffusion to smear out inhomogeneities on the length scale
$R$. Correspondingly, the Damk{\"o}hler number quantifies to which
extent a sizable inhomogeneity can be induced by the activity near the
particle against the opposing effect of diffusion. A small value
implies that the chemicals which are produced by the activity do not
``crowd'' near the particle (i.e., the kinetics of the activity is in
the so-called ``reaction-limited'' regime), and the perturbations
$\delta n$ will be small. The numerical estimate of $\mathrm{Da}$
requires knowledge of the reaction rate $q$: for example, the Pt
catalyzed decomposition of hydrogen peroxide reported in
Refs.~\cite{Paxton2004,Brown2017} leads to $\mathrm{Da}\sim 10^{-1}$.

Therefore, we consider the linearized chemical potential of the
chemical species, 
\begin{equation}
  \label{eq:linmu}
  \mu \approx \mu_0 +
  \lambda^2 \left[ \frac{\delta n}{\xi^2}
    - \nabla^2 \delta n \right] ,
\end{equation}
with the correlation length defined as
\begin{equation}
  \label{eq:xi}
  \xi := \frac{\lambda}{\sqrt{h''(n_0)}}.
\end{equation}
Both quantities $\lambda$ and $h''(n_0)$ depend on temperature. In
particular, $1/h''(n_0)$ represents a susceptibility, and a possible
critical point of the model for the chemical solute can be identified
by the condition $h''(n_0)=0$. The temperature dependence is
model--specific; for instance, $h(n)$ could be approximated by, e.g.,
the ideal gas model (as done in the main text for the purpose of
quantitative estimates), the van der Waals model, or a mean-field
Ginzburg--Landau model. However, the formalism presented in the
manuscript does not require the explicit form of $h(n)$ to be
specified, as only its general properties are used. It is actually
more advantageous to parametrize the state by $\xi$ explicitly instead
of by the temperature. Consequently, in our formulation the two
relevant independent parameters of the model for the chemical are
$\lambda$ and $\xi$, while the activity is parametrized by the
Damk{\"o}hler number $\mathrm{Da}$.

In this linear approximation, Eq.~(\ref{eq:chemstat}) reduces to
\begin{equation}
  \label{eq:laplacianmu}
  \nabla^2 \mu(\br) = 0.
\end{equation}
By choosing the origin of the coordinates to coincide with the center of the 
spherical particle, the boundary conditions can be written as (see 
Eqs.~(\ref{eq:bcinfty}) and (\ref{eq:bcR}) in the main text)
\begin{equation}
  \label{eq:bcmulin}
  \frac{\partial \mu}{\partial r}(\br=R\be_r)
  = -\frac{q}{n_0 \Gamma} \mathbb{A}(\be_r) ,
  \qquad
  \mu(\br\to\infty) \to \mu_0 
\end{equation}
for the chemical potential and (see Eqs.~(\ref{eq:bcinfty}) and
(\ref{eq:activity}) in the main text)
\begin{equation}
  \label{eq:bcnlin}
  \lambda^2 \frac{\partial \delta n}{\partial r}(\br=R\be_r) = 0 ,
  \qquad
  \delta n(\br\to\infty)\to 0 
\end{equation}
for the density of the chemical.

In order to proceed, one first solves the Laplace 
equation~(\ref{eq:laplacianmu}), with the activity field in \eq{eq:alm} 
entering through the boundary condition in \eq{eq:bcmulin}. This gives
\begin{equation}
  \label{eq:deltamu}
  \mu(\br) = \mu_0 +
  \frac{\mathrm{Da}}{\beta}
  \sum_{\ell m} 
  \frac{a_{\ell m}}{\ell +1} \,
  \left(\frac{R}{r}\right)^{\ell+1}\,
  Y_{\ell m}(\theta,\varphi) ,
\end{equation}
in terms of the spherical harmonics $Y_{\ell m}(\theta,\varphi)$ (see
\eq{eq:Ylm}). Next one solves the inhomogeneous 
Helmholtz equation~(\ref{eq:linmu}) for $\delta n(\br)$ with the chemical 
potential $\mu(\br)$ found above (\eq{eq:deltamu}). In view of 
\eq{eq:laplacianmu}, one can introduce the auxiliary dimensionless field
$\Psi(\br)$ via
\begin{equation}
  \label{eq:deltan}
  \delta n(\br)
    = \left(\frac{\xi}{\lambda}\right)^2 \left[ \mu(\br) - \mu_0
  \right]+ n_0 \Psi(\br) ,
\end{equation}
so as to obtain a homogeneous Helmholtz equation,
\begin{equation}
  \nabla^2 \Psi - \frac{\Psi}{\xi^2}
  = 0 ,
\end{equation}
together with the boundary conditions following from 
\eq{eq:bcnlin}:
\begin{equation}
  \Psi(\br) \to 0 ,
  \qquad
  \mathrm{for}\; |\br|\to \infty ,
\end{equation}
\begin{equation}
  \frac{\partial \Psi}{\partial r}(\br)
  = - \frac{1}{n_0} \left(\frac{\xi}{\lambda}\right)^2
  \frac{\partial \mu}{\partial r}(\br)
  = \frac{\mathrm{Da}}{R \beta n_0}
    \left(\frac{\xi}{\lambda}\right)^2
    \sum_{\ell m}  a_{\ell m} \, Y_{\ell m}(\theta,\varphi) ,
  \qquad
  \mathrm{for}\; |\br|=R.
\end{equation}
The solution of this boundary problem is given straightforwardly by
\begin{equation}
  \label{eq:Psi}
  \Psi(\br) = \frac{\mathrm{Da} \, R^2}{\beta n_0 \lambda^2}
  \sum_{\ell m} c_\ell \left(s=\frac{r}{R}, 
      \tilde{\xi} = \frac{\xi}{R} \right) \,
  a_{\ell m}\, Y_{\ell m}(\theta,\varphi) ,
\end{equation}
where we have introduced the coefficients
\begin{equation}
  \label{eq:ccoeff}
  c_\ell(s,\tilde{\xi}):= \psi_\ell(\tilde{\xi})
  \frac{K_{\ell+1/2}(s/\tilde{\xi})}{\sqrt{s}} 
  \end{equation}
and
\begin{equation}
  \label{eq:psiell}
  \psi_\ell(\tilde{\xi}) :=
  - \frac{\tilde{\xi}^{2}}{(\ell+1)
    K_{\ell+1/2}(\tilde{\xi}^{-1})
    + \tilde{\xi}^{-1} K_{\ell-1/2}(\tilde{\xi}^{-1})}
  \qquad
  (\leq 0) 
\end{equation}
in terms of the modified Bessel function of the second kind
$K_{p}(z)$
  \cite{abram,GrRy94}. For future reference, we also quote the
  explicit solution for the field $\delta n(\br)$ given by
  \eq{eq:deltan}:
\begin{equation}
  \delta n(\br) = \frac{\mathrm{Da} \, R^2}{\beta \lambda^2}
  \sum_{\ell m} \hat{c}_\ell \left(\frac{r}{R}, \frac{\xi}{R} \right) \,
  a_{\ell m} \, Y_{\ell m}(\theta,\varphi) ,
\end{equation}
with the coefficients
\begin{equation}
  \label{eq:hatccoeff}
  \hat{c}_\ell(s,\tilde{\xi}) :=
  c_\ell(s,\tilde{\xi})
  + \frac{\tilde{\xi}^2}{(\ell+1) s^{\ell+1}} .
\end{equation}

As anticipated, the typical magnitude of the perturbation in the
chemical potential is set by the Damk{\"o}hler number (see
\eq{eq:deltamu}) whereas the magnitude of the field $\Psi$ (see
\eq{eq:Psi}) is also fixed by $\mathrm{Da}$ but it is modulated by the
ratio of the particle size and a characteristic length
$\sqrt{\beta n_0 \lambda^2}$ associated with the term
$\propto \lambda^2$ in the chemical potential (see
\eq{eq:mu}). (If the local free energy is approximated by the
  ideal--gas form, $\beta h(n)=n (\ln n-1)$, this characteristic
  length is actually the correlation length defined by \eq{eq:xi}.)

As a final remark, we note that those changes in the boundary
condition for $\delta n(\br)$ at the surface of the particle which
preserve its homogeneous character, do not alter the structure of the
solution given by \eq{eq:Psi}, but only the value of the factors
$\psi_\ell(\tilde{\xi})$ given by \eq{eq:psiell}. For instance, a
Dirichlet boundary condition (which is the other option ensuring a
vanishing surface term in the functional derivative in \eq{eq:mu}),
i.e.,
\begin{equation}
  \delta n(\br) = 0,
  \qquad
  \mathrm{for}\; |\br|=R,
\end{equation}
would translate into the boundary condition
\begin{equation}
  \Psi(\br)
  = -\frac{1}{n_0} \left(\frac{\xi}{\lambda}\right)^2
  \left[ \mu(\br) - \mu_0 \right]
  = -\frac{\mathrm{Da}}{\beta n_0}
    \left(\frac{\xi}{\lambda}\right)^2
  \sum_{\ell m} 
  \frac{a_{\ell m}}{\ell +1} \,
  Y_{\ell m}(\theta,\varphi) ,
  \qquad
  \mathrm{for}\; |\br|=R.
\end{equation}
This leads to a solution like in \eq{eq:Psi}, but with the factors
\begin{equation}
  \psi_\ell(\tilde{\xi}) \mapsto  \psi_\ell^{(D)}(\tilde{\xi}) = -
  \frac{\tilde{\xi}^2}{(\ell +1) K_{\ell+1/2}(\tilde{\xi}^{-1})} 
\end{equation}
in the coefficients defined by Eq.~(\ref{eq:ccoeff}).

\section{The hydrodynamical problem}
\label{sec:hydr}

The hydrodynamical problem for the velocity field $\bu(\br)$ is given
by Eqs.~(\ref{eq:stokes}, \ref{eq:f}) with the boundary conditions
given by Eqs.~(\ref{eq:bcinfty}, \ref{eq:noslip}) and with the fields
$n(\br)$ and $\mu(\br)$ obtained as solutions of the diffusion problem
discussed in Sec.~\ref{sec:chem}. The assumption of slow particle
motion allows one to neglect the effects of compressibility (i.e.,
small Mach number), convection (i.e., small Reynolds number), and
time--dependence on the flow. For small Mach and Reynolds numbers, the
flow dynamics reduces to the diffusion of vorticity (time--dependent
Stokes equation \cite{BrennerBook}), which is characterized by the
diffusion coefficient $D_\mathrm{vort} = \eta/\rho$, in terms of the
viscosity $\eta$ and the mass density $\rho$ of the fluid. Thus, one
only needs to repeat the reasoning presented in Sec.~\ref{sec:chem}
for the chemical diffusion, but with the replacement
$D_\mathrm{chem} \to D_\mathrm{vort}$. For water one has
$\eta = 10^{-3}\,\mathrm{Pa\, s}$ and $\rho = 10^3\,\mathrm{kg/m^3}$,
resulting in a diffusion coefficient
$D_\mathrm{vort} = 10^{-6}\, \mathrm{m^2/s}$, i.e., three orders of
magnitude larger than the estimate of $D_\mathrm{chem}$. One can
therefore conclude that, concerning the hydrodynamical flow, the
slow--particle approximation is even more justified.

The ultimate goal of solving the hydrodynamical problem for the field
$\bu(\br)$ is to obtain the velocities of translation, $\bV$, and of
rotation, $\bOmega$, of the particle. There is, however, an approach,
based on the generalized Lorentz reciprocal theorem
\cite{Lorentz_original,Lorentz_transl,Teubner1982,KiKa91}, which
allows one to sidestep the explicit computation of the hydrodynamic
flow field $\bu(\br)$, and instead to express $\bV$ and $\bOmega$
explicitly in terms of the body force $\mathbf{f}$ only.

\subsection{The generalized Lorentz reciprocal theorem} 

This theorem states the following relation between any two Newtonian fluid flows
$\{\eta,\mathbf{u}, \Pitens\}$ and
$\{\eta',\mathbf{u}',\Pitens'\}$ (the notation refers to
\{viscosity, flow field, hydrodynamic stress tensor\}) which are solutions, 
in the same domain ${\cal D}$, of the incompressible Stokes equations
(Eq.~(\ref{eq:stokes})) with (sufficiently well behaved) force
densities $\mathbf{f}$ and $\mathbf{f}'$:
\begin{equation}
 \label{eq:Lorentz_th}
\eta' 
\left[\,\,\oint \limits_{\partial {\cal D}} \mathbf{u}' \cdot 
\Pitens \cdot d\boldsymbol{\mathcal{S}}
- \int\limits_{{\cal D}} 
\vol\;
\mathbf{u}' \cdot \mathbf{f} 
\right] 
= 
\eta \left[\,\,\oint \limits_{\partial {\cal D}} \mathbf{u} \cdot 
\Pitens' \cdot d\boldsymbol{\mathcal{S}}  -
\int\limits_{{\cal D}} 
\vol\;
\mathbf{u} \cdot \mathbf{f}' 
\right]\,.
\end{equation}
Here the orientation of the surface element
$d\boldsymbol{\mathcal{S}} := \be_n\; \surf$ is chosen to be the one
given by the \textit{inner} normal (i.e., pointing \textit{into} the
fluid domain ${\cal D}$), in agreement with the convention used in
Sec.~\ref{sec:forceBalance}. For the present system, the fluid domain
${\cal D}$ is the volume of the solution outside the particle; thus
$\partial {\cal D}$ is composed of the surface
$\partial {\cal D}_{part}$ of the particle and a distant (at infinity)
surface $\partial{\cal D}_\infty$. The hydrodynamic stress tensor is
related to the velocity field as follows:
\begin{equation}
  \label{eq:pres_tens}
  \Pitens = \eta \left[ \nabla\bu 
    + (\nabla\bu)^\dagger \right] -
  \Itens p .
\end{equation}

For convenience, a brief derivation of the theorem is included here (see the 
textbook by Kim and Karrila \cite{KiKa91}). By applying Gauss' theorem 
\cite{BrennerBook} for the (sufficiently well behaved, but 
otherwise arbitrary) tensor and vector fields $\Pitens$, 
$\mathbf{u}$, and $\mathbf{f}$, as well as for the primed fields, 
respectively, one can re-write the left-hand side of Eq.~(\ref{eq:Lorentz_th}) 
as (note the use of Einstein's
convention of summation over repeated indices and of the
\textit{inner} normal $\be_n$ for the oriented surface element):
\begin{eqnarray}
\label{proof_rec_th}
{\cal A} & := & \eta' \left[\,\int \limits_{\partial {\cal D}} \mathbf{u}' 
\cdot \Pitens \cdot d\boldsymbol{\mathcal{S}}
- \int\limits_{{\cal D}} 
\vol\;
\mathbf{u}' \cdot \mathbf{f} 
\right] 
 = 
\eta' \int\limits_{{\cal D}} 
\vol\;
\left[ - \nabla \cdot \left(\mathbf{u}' 
\cdot \Pitens\right) - \mathbf{u}' \cdot \mathbf{f} 
\,\right]
\nonumber\\
& = & \eta' \,\int \limits_{{\cal D}}
\vol\;
\left[ -(\partial_j u'_i) \mathsf{\Pi}_{ij} 
- \mathbf{u}' \cdot \left(\nabla \cdot \Pitens + \mathbf{f} 
\right) 
\right] 
\ \stackrel{\mathrm{Eq.\,(\ref{eq:stokes})}}{=} 
-\eta' \int \limits_{{\cal D}}
\vol\; 
(\partial_j u'_i) \mathsf{\Pi}_{ij} 
\nonumber\\
& \stackrel{i \leftrightarrow j}{=} & 
-\frac{1}{2} \,\eta' \int \limits_{{\cal D}}
\vol\; 
\left[(\partial_j 
u'_i) \mathsf{\Pi}_{ij} 
+ 
(\partial_i u'_j) \mathsf{\Pi}_{ji} \right]
\stackrel{\mathsf{\Pi}_{ij} = \mathsf{\Pi}_{ji}}{=} 
- \frac{1}{2} \int \limits_{{\cal D}} 
\vol\;
\left[\eta' (\partial_j u'_i + 
\partial_i u'_j) \right] \,\mathsf{\Pi}_{ij}
\nonumber\\
&\stackrel{\mathrm{Eq.\,(\ref{eq:pres_tens})}}{=}& 
- \frac{1}{2} \int \limits_{{\cal D}} 
\vol\;
\mathsf{\Pi}'_{ij} \mathsf{\Pi}_{ij} \,d^3 
\mathbf{r} 
- \frac{1}{2} \int \limits_{{\cal D}} 
\vol\; p' 
\delta_{ij} \mathsf{\Pi}_{ij}
\nonumber\\
&\stackrel{\mathrm{Eq.\,(\ref{eq:pres_tens})}}{=}&
-  \frac{1}{2} \int \limits_{{\cal D}} 
\vol\;
\Pitens : \Pitens 
-   \eta \int \limits_{{\cal D}} 
\vol\;
p' \, (\nabla \cdot \mathbf{u})
+ \frac{3}{2} \int \limits_{{\cal D}} 
\vol\;
p' p 
\nonumber\\
& \stackrel{\mathrm{Eq.\,(\ref{eq:stokes})}}{=} &
\frac{1}{2} \int \limits_{{\cal D}} 
\vol\;
\left[- \Pitens' : \Pitens + 3 p' p 
\right]
\,.
\end{eqnarray}
A similar sequence of transformations can be applied to the right-hand
side of \eq{eq:Lorentz_th} (because the fields involved are defined in
the same domain), rendering as a result the last line above (up to an
irrelevant swapping of the primed and unprimed quantities). Accordingly, one 
concludes that both sides of \eq{eq:Lorentz_th} are equal.

\subsection{Translational and rotational velocities}

We employ the Lorentz theorem by choosing for the unprimed system the fluid 
motion driven by the body force distribution $\mathbf{f}(\mathbf{r})$ and 
corresponding to presence of the force- and torque-free self-propelled 
particle, which translates with velocity $\mathbf{V}$ and rotates with 
angular velocity $\bOmega$ (with respect to the center of mass) 
through an otherwise quiescent fluid. For the primed system, the choice is
that of the flow of a fluid of the same viscosity (i.e., $\eta' = \eta$), which 
is induced by a no-slip particle of the same shape moving with constant 
velocities $\mathbf{V}'$ and $\bOmega'$ (through an otherwise 
quiescent fluid) due to being acted upon by an external force $\mathbf{F}'$ and 
an external torque $\boldsymbol{\tau}'$, in the absence of body forces acting 
on the fluid, i.e., $\mathbf{f}' \equiv 0$.

\subsubsection{Particle of arbitrary shape}
By taking the origin of the system of coordinates to coincide with the center of 
mass of the particle, the corresponding boundary conditions turn into 
\begin{equation}
  \label{eq:St_sol_1}
  \mathbf{u}(\mathbf{r} \in {\partial\cal D}_{part}) = \mathbf{V} + 
  \bOmega \times \mathbf{r} \,, 
~~\mathbf{u}(|\br| \to 
  \infty) = 0\,,
  \qquad\textrm{[see Eqs.~(\ref{eq:bcinfty}, \ref{eq:noslip})]}
\end{equation}
and similarly for the primed system. The field $\mathbf{u}'$ decays as $1/r$, 
corresponding to the Stokeslet induced by the localized force $\mathbf{F}'$ 
pulling the particle. The field $\mathbf{u}$ can also be assumed to decay 
at least as $1/r$, provided that the external potential $W(\br)$, and thus 
the body force $\mathbf{f}(\br)$ (see \eq{eq:fexpand}), decays sufficiently 
fast (for a finite correlation length, the specific contribution to 
$\mathbf{f}(\br)$ by the chemical decays exponentially, see 
Sec.~\ref{sec:chem}). Therefore, the surface integrals over $\partial{\cal 
D}_\infty$ vanish, and in Eq.~(\ref{eq:Lorentz_th}) only the integrals 
over the surface $\partial{\cal D}_{part}$ of the particle contribute. One can
insert the expressions for the flows $\mathbf{u}$ and $\mathbf{u}'$
from the boundary conditions (\eq{eq:St_sol_1}) into the surface
integrals over the surface of the particle. The expression can be simplified by 
noting the following points: 
\begin{enumerate}
\item[(i)] $\mathbf{V}$, $\bOmega$, $\mathbf{V}'$, and
  $\bOmega'$ are constant vectors over the surface of the
  particle.
\item[(ii)]
  $(\bOmega' \times \mathbf{r}) \cdot (\Pitens
  \cdot \be_n) \equiv \bOmega' \cdot [\mathbf{r} \times
  (\Pitens \cdot \be_n)]$ (and similarly for
  $\bOmega$ and $\Pitens'$).
\item[(iii)] The integrals of $\Pitens' \cdot \be_n$ and
  $\mathbf{r} \times (\Pitens' \cdot \be_n)$ over the
  surface of the particle represent the force and torque,
  respectively, by the fluid on the particle (thus they are equal in
  magnitude and opposite in sign to the external force $\mathbf{F}'$
  and torque $\boldsymbol{\tau}'$, respectively). 
\item[(iv)] Due to Gauss' theorem, the integral of
  $\Pitens \cdot \be_n$ over the surface of the particle is equal to
  the volume integral of $\mathbf{f}(\mathbf{r})$ (see
  Eqs.~(\ref{eq:integralstokes}) and (\ref{eq:nohydroshear})), and
  similarly the integral of $\mathbf{r} \times (\Pitens \cdot \be_n)$
  over the surface of the particle can be written as a volume integral
  of $\br\times\mathbf{f}(\mathbf{r})$ (see Eqs.~(\ref{eq:torquepart})
  and (\ref{eq:nohydrotorque})). This is the step at which the
  mechanical isolation of the system is imposed upon solving the
  hydrodynamical problem. If the particle would experience a force or
  torque from an external source, there would be a nonzero
  contribution from the integrated stresses at infinity.
\end{enumerate}
In this manner, one arrives at
\begin{equation}
  \label{eq:Lorentz_interm_1}
   \mathbf{V}' \cdot \int\limits_{{\cal D}} 
   \vol\;
   \mathbf{f} 
   + \bOmega' \cdot \int\limits_{{\cal D}}
   \vol\;
   \br\times\mathbf{f} 
   - \int\limits_{{\cal D}} 
   \vol\;
   \mathbf{u}' \cdot \mathbf{f} 
   = - \mathbf{V} \cdot \mathbf{F}' -
\bOmega \cdot \boldsymbol{\tau}'\,.
\end{equation}
Due to the linearity of the Stokes equations, the flow $\mathbf{u}'(\br)$ 
(i.e., the solution of the incompressible Stokes equations without body 
forces or torques) depends linearly on the boundary condition at the 
surface of the particle, i.e., on $\mathbf{V}'$ and $\bOmega'$, and 
thus can be written as \cite{BrennerBook}
\begin{equation}
  \label{eq:u_prime_from_bc}
  \mathbf{u}'(\mathbf{r}) =
  \bV'\cdot\Ttens(\br) 
  + \bOmega'\cdot\Ktens(\br) ,
\end{equation}
where the second--rank tensors $\Ttens(\br)$ 
and $\Ktens(\br)$ depend \textit{solely} on the shape of the particle 
and on the boundary conditions, and can thus be considered to be known. 
Likewise, the forces and velocities are also related linearly. This can be 
expressed in a compact manner by introducing the supervector notation
\begin{equation}
 \label{eq:gen_vel_forc}
 \mathbb{V} :=  \left(
 \begin{array}[c]{c}
      \mathbf{V}\\
     \bOmega
    \end{array} 
  \right)\,,~~
  \qquad
  \mathbb{F} :=  \left(
 \begin{array}[c]{c}
      \mathbf{F}\\
     \boldsymbol{\tau}
    \end{array} 
    \right)\,,   
\end{equation}
(and likewise for the primed system). This is a formal procedure in the 
sense that the components do not have the same units, and both true vectors 
(velocity, force) and pseudovectors (angular velocity, torque) are mixed. This 
allows one to write \cite{BrennerBook}
\begin{equation}
 \label{eq:res_tens}
 \mathbb{F} = {\mathbf{\cal R}} \cdot  \mathbb{V}
 \qquad \Leftrightarrow \qquad 
 \mathbb{V} = {\mathbf{\cal M}} \cdot \mathbb{F} 
\end{equation}
in terms of the resistance supertensor ${\mathbf{\cal R}}$ and its
inverse, the mobility supertensor ${\mathbf{\cal M}}$, which both
also depend only on the shape of the particle and the boundary
conditions.

By inserting \eq{eq:u_prime_from_bc} into \eq{eq:Lorentz_interm_1} and
defining the supervector
\begin{equation}
  \mathbb{T}(\br) :=
  \left(
    \begin{array}[c]{c}
      \left[ \Ttens(\br) - \Itens
      \right]\cdot\mathbf{f}(\br)
      \\
      \Ktens(\br)\cdot\mathbf{f}(\br) - \br\times\mathbf{f}(\br)
    \end{array}
  \right) ,
\end{equation}
the Lorentz theorem takes the compact form
\begin{equation}
  \mathbb{V}\cdot\mathbb{F}'
  = \left[
    \int\limits_{{\cal D}}
    \vol\;
    \mathbb{T}(\br)
  \right] \cdot \mathbb{V}'
  = \left[
    \int\limits_{{\cal D}} 
    \vol\;
    \mathbb{T}(\br)
  \right] \cdot {\mathcal{M}}\cdot\mathbb{F}' ,
\end{equation}
where the second equality follows from \eq{eq:res_tens}. Since this 
relation holds for arbitrary choices of the force $\mathbf{F}'$ and of 
the torque $\boldsymbol{\tau}'$ in the primed problem, one straightforwardly 
obtains the translational and rotational velocities of the active particle as
\begin{equation}
    \label{eq:U_and_V_general}
    \mathbb{V} =
    \int\limits_{{\cal D}}
    \vol\;
    \mathbb{T}(\br) \cdot {\mathbf{\cal M}}\;.
  \end{equation}

\subsubsection{The case of a spherical particle}
We now particularize the expression above for the case of a sphere
of radius $R$, for which the rotation and the translation are
decoupled. In this case one has \cite{BrennerBook} 
\begin{equation}
  {\mathbf{\cal M}} =
  \left(
    \begin{array}[c]{cc}
      \displaystyle \frac{\Itens}{6 \pi \eta R} 
      & \mathsf{0}
      \\
      \\
      \mathsf{0}
      & \displaystyle \frac{\Itens}{8 \pi \eta R^3}
    \end{array}
  \right)
\end{equation}
for the mobility supertensor. The tensors $\Ttens(\br)$ and
$\Ktens(\br)$, which relate the flow at $\br = r \be_r$
(within the domain occupied by the fluid, i.e., $r > R$) with the
boundary condition on the surface of the spherical particle (see
\eq{eq:u_prime_from_bc}), are given by the expressions
\cite{BrennerBook}
\begin{equation}
  \Ttens(\br)
  := 
  \frac{3R}{4} \left[ \Otens(\br) + \frac{R^2}{6} \nabla^2
    \Otens(\br)\right]
  = 
  \frac{3 R}{4 r}\left(1 +\frac{R^2}{3 r^2} \right) 
  \Itens + 
  \frac{3 R}{4 r}\left(1 - \frac{R^2}{r^2} \right) \be_r \be_r ,
  \label{eq:Ttens}
\end{equation}
where 
\begin{equation}
  \label{eq:oseen}
  \Otens(\br) = \frac{1}{r} \left[ \Itens +
    \be_r \be_r \right]
\end{equation}
is the Oseen tensor (i.e., Green's function for the unbounded Stokes 
equation), and
\begin{equation}
  \Ktens(\br) := - \left(\frac{R}{r}\right)^3 r \, \be_r
  \cdot \mathsf{\epsilon} \, ,
\end{equation}
in terms of the tensor $\mathsf{\epsilon} = -
\Itens\times \Itens = \be_i\be_j\be_k 
\varepsilon_{ijk}$ involving the three-dimensional permutation 
(Levi-Civita) tensor. (We note that the constraint of incompressible
flow implies $\nabla\cdot\Otens=0$, $\nabla\cdot\Ttens=0$, and 
$\nabla\cdot\Ktens=0$.)
Accordingly, from Eq.~(\ref{eq:U_and_V_general}) it follows that
\begin{equation}
 \label{velocity_self_propel}
 \mathbf{V} = \frac{1}{6 \pi \eta R} \int\limits_{r>R} 
 \vol\;
 \left[ \Ttens(\br) -  \Itens \right] \cdot  
 \mathbf{f}(\mathbf{r}) 
\end{equation}
and
\begin{equation}
  \label{rot_self_propel} 
  \bOmega = \frac{1}{8 \pi \eta R^3} \int\limits_{r>R} 
  \vol\;
  \left[
    \Ktens(\br)\cdot\mathbf{f}(\br) -
    \br\times\mathbf{f}(\br)
  \right]
  = \frac{1}{8 \pi \eta R^3} \int\limits_{r>R} 
  \vol\;
  \left[ \left(\frac{R}{r}\right)^3 - 1 \right] r\,
  \be_r \times \mathbf{f}(\mathbf{r})\,.
\end{equation}

\subsubsection{Alternative expressions for the drift and rotational
  velocities}

There are alternative ways of expressing
Eqs.~(\ref{velocity_self_propel}) and (\ref{rot_self_propel}) which
provide further physical insight. As remarked above, the source of the
motion is the curl of the force field $\mathbf{f}$, and indeed it is
possible to express $\mathbf{V}$ and $\bOmega$ as explicit functionals
of $\nabla\times\mathbf{f}$. Since the kernel in the integral in
\eq{rot_self_propel} is a potential field, one can establish the
following relations:
\begin{equation}
  \left[ \left(\frac{R}{r}\right)^3 - 1 \right] r\, \be_r 
  = - \nabla Q(r) ,
  \qquad
  \mathrm{with}\;\;
  Q(r) = - \frac{R^2}{10} \left[ 3 - 5 \left (\frac{r}{R}\right)^2
    + 2 \left (\frac{r}{R}\right)^5 \right] ,
\end{equation}
and \eq{rot_self_propel} can be recast as
\begin{equation}
  \label{eq:Omegacurlf}
  \bOmega = \frac{1}{8 \pi \eta R^3} \int\limits_{r>R} 
  \vol\;
  Q(r) \, \nabla \times \mathbf{f}(\mathbf{r})\, ,
\end{equation}
after integration by parts for a field $\mathbf{f}$ that decays
sufficiently fast at infinity. The contribution due to the boundary
integral at $r=R$ vanishes, because $Q(r=R)=0$ by construction. In a
similar fashion, the incompressibility constraint
$\partial_j (\mathsf{T}_{ij}-\delta_{ij})=0$ implies the existence,
for each value of the index $i$, of a vector potential field
$\mathbf{A}_i$ in the domain $r>R$, which can be obtained by solving
three boundary-value problems (one for each value of the index $i$):
\begin{equation}
  \nabla \times \mathbf{A}_i = \be_i \cdot \left( \Ttens -
    \Itens \right), 
  \qquad
  \nabla \cdot \mathbf{A}_i = 0 ,
  \qquad
  \be_r \times \mathbf{A}_i = 0 
  \;
  \mathrm{at}\; |\br| = R ,
  \qquad
  \mathbf{A}_i \to \frac{\br\times\be_i}{2} 
  \;
  \mathrm{as}
  \; |\br|\to \infty .
\end{equation}
The choice of gauge ($\nabla\cdot\mathbf{A}_i=0$) and of the boundary
condition for the tangential component at $r=R$ is arbitrary; the
boundary condition at infinity is determined by the inhomogeneous term
$\Itens$. The solution of this problem\footnote{This follows from
  expanding the fields in terms of spherical harmonics and by noting
  that
  $\nabla\times(\be_i\cdot\Ttens) = (3R/4)
  \nabla\times(\be_i\cdot\Otens) = -(3R/2) \be_i\times\nabla(1/r)$
  (see \eq{eq:oseen}). The solution is determined up to the gradient
  of a harmonic function.} is given by
\begin{equation}
  \label{eq:potA}
  \mathbf{A}_i(\br) = \frac{3 R}{4} \left[ 1 - \frac{2r}{3R} -
    \frac{R^2}{3r^2} \right] \be_i\times\be_r .
\end{equation}
Consequently, \eq{velocity_self_propel} can be rearranged also as
\begin{equation}
  \label{eq:Vcurlf}
  V_i = \frac{1}{6 \pi \eta R} \int\limits_{r>R} 
  \vol\;
  [\nabla\times\mathbf{A}_i (\br)] \cdot \mathbf{f}(\br)
  =
  \frac{1}{6 \pi \eta R} \int\limits_{r>R} 
  \vol\;
  \mathbf{A}_i \cdot \left(\nabla\times\mathbf{f}\right) ,
\end{equation}
after integration by parts for a field $\mathbf{f}$ that decays
sufficiently fast at infinity (the contribution by the boundary
integral at $r=R$ vanishes because $\mathbf{A}_i(\br=R\be_r)=0$ by
construction).

Further simplification is possible in the particular case that the
force is due solely to the chemical stress tensor (see
\eq{eq:Pichem}),
\begin{equation}
  \mathbf{f} = \nabla\cdot\Pitens_\mathrm{chem},
  \qquad    
  \Pitens_\mathrm{chem}
  = \lambda^2\left[ \frac{1}{2} \nabla\nabla (n^2) - (\nabla n)
    (\nabla n) \right]
  - \Itens \, \left[
    \frac{\lambda^2}{2} |\nabla n|^2 + p_\mathrm{chem}(n)
  \right] .
\end{equation}
In this manner, \eq{velocity_self_propel} turns into
\begin{equation}
  \label{eq:symetricV}
  \mathbf{V} = - \frac{1}{6 \pi \eta R} \int\limits_{r>R} \vol \;
  \Pitens_\mathrm{chem} : \nabla\Ttens 
  = \frac{\lambda^2}{6 \pi \eta R} \int\limits_{r>R} \vol \;
  (\nabla n) (\nabla n) : \nabla\Ttens ,
\end{equation}
after integration by parts and use of the incompressibility
constraint. (We note that
$\nabla\Ttens (\br=R \be_r) = (3/2R) \be_r (\be_r \be_r - \Itens)$
(see \eq{eq:Tsmallr}), so that the contribution by the term
$\propto\nabla\nabla (n^2)$ to the boundary integral at $r=R$ does
vanish.) Likewise, \eq{rot_self_propel} can be written as (summation
over repeated indices is assumed)
\begin{equation}
  \label{eq:symetricOmega}
  \bOmega = \frac{1}{8 \pi \eta R^3} \int\limits_{r>R} 
  \vol\;
  \be_i \, \varepsilon_{ijk}
  \left(\Pitens_\mathrm{chem}\right)_{jm}
  (\be_k\cdot \nabla)
  (-\be_m\cdot \nabla Q)
  = \frac{3\lambda^2}{8 \pi \eta} \int\limits_{r>R} 
  \vol\;
  \frac{1}{r^3} \, \be_r \cdot (\nabla n) (\nabla n) \times 
  \be_r.
\end{equation}
Here the term $\propto\nabla\nabla (n^2)$ does not contribute because
the operator $\be_r\times \nabla$ is essentially the derivative
tangential to spheres $r=\mathrm{constant}$, so that the angular
integral over the whole spherical surface renders zero.
Equations~(\ref{eq:symetricV}) and (\ref{eq:symetricOmega}) show
explicitly the symmetric manner in which the inhomogeneity induced by
the activity enters the expressions for $\bV$ and $\bOmega$ (compare
also with \eq{eq:curlf}).

\subsubsection{Fax{\'e}n's law}

Finally, one can derive a suggestive interpretation of the velocities
of translation and rotation. To this end, we introduce the velocity
field
\begin{equation}
  \label{eq:uamb}
  \mathbf{u}^\infty(\mathbf{r}) = \frac{1}{8\pi\eta}
  \int\limits_{r' > R}
  \vol\;
  \Otens(\mathbf{r}'-\mathbf{r}) \cdot
  \mathbf{f}(\mathbf{r}') \,,
\end{equation}
in terms of the Oseen tensor (Eq.~(\ref{eq:oseen})). Physically, 
$\mathbf{u}^\infty(\br)$ is the velocity field that would be induced by the 
chemical bulk force $\mathbf{f}$ \emph{alone}, i.e., without regard to the 
boundary conditions imposed by the particle. (In other 
words, $\mathbf{u}^\infty(\mathbf{r})$ is the flow in an unbounded fluid, as 
if the particle would not be present; correspondingly, it is also 
defined in the region $|\br|<R$.) In this sense, $\mathbf{u}^\infty(\mathbf{r})$ 
plays the role of an \emph{ambient flow} for the particle. If 
\eq{velocity_self_propel} is evaluated by using the definition in \eq{eq:Ttens} 
and the balance of chemical forces (\eq{eq:fchembalance}), one arrives at
\begin{equation}
  \label{eq:faxen2}
  \mathbf{V} =
  \left[
    \mathbf{u}^\infty(\mathbf{r}) + \frac{R^2}{6} \nabla^2
    \mathbf{u}^\infty(\mathbf{r}) \right]_{\mathbf{r}=0}
  + \frac{1}{6\pi \eta R}\mathbf{F}^\mathrm{(part)}_\mathrm{chem}\,;
\end{equation}
i.e., the velocity of the particle is the sum of a drag by the ambient
flow and of a direct pull on the particle by the interaction with the
molecules of the chemical. This expression can be recast in
the form
\begin{equation}
  \label{eq:faxen}
  6\pi \eta R \mathbf{V} - 6\pi\eta R \left[
    \mathbf{u}^\infty(\mathbf{r}) + \frac{R^2}{6} \nabla^2
    \mathbf{u}^\infty(\mathbf{r}) \right]_{\mathbf{r}=0}
  = \mathbf{F}^\mathrm{(part)}_\mathrm{chem}.
\end{equation}
This is Fax\'en's law \cite{KiKa91}, which simply expresses the force
balance for the particle: the hydrodynamical force exerted by the
fluid on the particle (left hand side of the equation) is equal to the
resultant of any other forces (right hand side). From this
perspective, \eq{eq:faxen} is just a disguised form of
Eqs.~(\ref{eq:Fpart}) and (\ref{eq:Fpartzero}), but with an explicit
expression for the integral of the hydrodynamic stress $\Pitens$ over
the surface of the particle. Likewise, there is also Fax\'en's law for
the torque:
\begin{equation}
  8\pi\eta R^3 \left[ \bOmega - \frac{1}{2}
    \nabla\times\bu^\infty(\br=0) \right]
  = \boldsymbol{\tau}^\mathrm{(part)}_\mathrm{chem}.
\end{equation}

\subsection{Velocity of a spherical particle as a function of
  activity}

By inserting the definition of $\mathbf{f}$ (\eq{eq:f} in the main 
text) into \eq{velocity_self_propel} one arrives at
\begin{equation}
  \label{eq:V}
  \bV = \frac{1}{6\pi\eta R} \int_{r>R}
  \vol\;
  \left[ \Ttens(\br) - \Itens \right] \cdot
  \left[ - \delta n(\br) \, \nabla \mu(\br) \right] ,
\end{equation}
after disregarding the gradient term $n_0 \nabla \mu$ which does not
contribute because it can be absorbed in the dynamic
pressure. (Alternatively, integration by parts in \eq{eq:V} gives zero
due to the incompressibility constraint $\nabla\cdot\Ttens=0$.) The
expression in \eq{eq:V} actually does not rely on any specific
mechanism of phoresis. (It follows solely from the Stokes equation and
the local equilibrium hypothesis embodied in \eq{eq:f}.) It shows
explicitly that a nonvanishing velocity $\bV$ requires both a
deviation from equilibrium, so that $\nabla\mu\neq 0$, and a deviation
from a homogeneous state, so that $\delta n\neq 0$ (compare with
\eq{eq:curlf}). In the model of correlation--induced transport we are
considering here, both requirements are enforced by the activity. This
explains why the final result (given by \eq{eq:tildeV} in the main
text) is bilinear in the activity. In the scenario of standard
phoresis, the activity is only needed to impose a deviation from
equilibrium, because the extended interaction $W$ is usually already
sufficient to bring about the necessary inhomogeneity (see
\eq{eq:driftW}, linear in the activity, as well as the comparative
analysis in Sec.~\ref{sec:W}).

For the particular model of correlation--induced transport, one can
insert the relationship given by \eq{eq:deltan} into \eq{eq:V} and neglect a 
perfect gradient term $\propto \nabla(\mu-\mu_0)^2$, so that the transport 
velocity can be written as
\begin{equation}
  \label{eq:Vpsimu}
  \bV = \frac{1}{6\pi\eta R} \int_{r<R} 
  \vol\;
  \left[ \Ttens(\br) - \Itens \right] \cdot \left[
    - n_0 \Psi(\br) \, \nabla \mu(\br) \right].
\end{equation}
One can split the integrand into tangential and radial components as
follows. The driving force is written as
\begin{equation}
  \label{eq:fradpar}
 - n_0 \Psi(\br) \, \nabla \mu(\br) = - n_0 \Psi \left[ 
    \frac{1}{r} 
    \tilde{\nabla}_\parallel \mu + \be_r
    \frac{\partial\mu}{\partial r} 
  \right] ,
\end{equation}
in terms of the gradient operator defined on the unit sphere:
\begin{equation}
  \label{eq:nablapar1}
  \tilde{\nabla}_\parallel := \be_\theta \frac{\partial}{\partial\theta} 
  + \frac{\be_\varphi}{\sin\theta} \frac{\partial}{\partial\varphi}\, .
\end{equation}
The hydrodynamic tensor multiplying the driving force is given by (see 
\eq{eq:Ttens})
\begin{equation}
  \Ttens(\br) -\Itens 
  = \left( \frac{3 R}{4 r} +\frac{R^3}{4 r^3} - 1 \right)
  \left( \Itens - \be_r \be_r \right) 
  + \left( \frac{3 R}{2 r} - \frac{R^3}{2 r^3} - 1 \right) \be_r \be_r ,
\end{equation}
where in the first summand one recognizes the tangential projector
$\Itens - \be_r \be_r$ onto the spherical surfaces centered at the 
origin, while the second summand involves the radial projector $\be_r \be_r$ 
along the normal to these surfaces. Therefore, one can write \eq{eq:Vpsimu} as
  \begin{equation} 
    \bV = -\frac{n_0}{6\pi\eta R} \int_{r<R} 
    \vol\;
    \Psi(\br) \left\{ \frac{1}{r} \left( \frac{3 R}{4 r} +\frac{R^3}{4
          r^3} - 1 \right) \tilde{\nabla}_\parallel \mu(\br) + \left(
        \frac{3 R}{2 r} - \frac{R^3}{2 r^3} - 1 \right) \be_r
      \frac{\partial\mu}{\partial r}(\br) \right\} .
\end{equation}
Introducing the dimensionless position vector $\bs := \br/R = (r/R) 
\be_r(\theta,\varphi)$ and the element of solid
angle $d\Omega := \sin\theta\;d\theta\,d\varphi$, we find
\begin{eqnarray}
  \bV
  & = 
  & -\frac{n_0 R}{6\pi\eta} \int_{s<1} ds\; s^2 \int d\Omega \; 
    \Psi(R s \be_r) 
    \begin{array}[t]{l}
      \displaystyle \left\{ 
      \frac{1}{s} \left( \frac{3}{4 s} +\frac{1}{4 s^3} - 1 \right)
      \tilde{\nabla}_\parallel \mu(R s \be_r) \right.
      \\
      \\
      \displaystyle \left . + \left( \frac{3}{2 s} - \frac{1}{2 s^3} - 1 \right) 
      \be_r \frac{\partial\mu}{\partial s}(R s \be_r) 
      \right\}  .
    \end{array}
\end{eqnarray}
The integral over the solid angle represents the scalar product in the
Hilbert space of the functions defined on the unit sphere, with
respect to which the spherical harmonics form an orthonormal
basis. Therefore, if the expansions in \eqs{eq:deltamu} and
(\ref{eq:Psi}) in terms of spherical harmonics for $\mu$ and $\Psi$
are inserted, one can exchange the angular integral with the sums over
the indices $\ell$ and $m$. Rearranging terms in this manner, one
arrives at the expression (see \eq{eq:tildeV} in the main text)
\begin{equation}
  \label{eq:driftV}
  \bV = V_0 \,
  \sum_{\ell m} \sum_{\ell' m'} 
  a_{\ell m} a_{\ell' m'} \, 
  \left[ 
    g^{\perp}_{\ell \ell'} \left(\frac{\xi}{R} \right) \,
    \bG^{\perp}_{\ell m; \ell' m'} 
    + 
    g^{\parallel}_{\ell \ell'} \left(\frac{\xi}{R} \right) \,
    \bG^{\parallel}_{\ell m; \ell' m'}
  \right] ,
\end{equation}
with the following definitions:
\begin{equation}
  \label{eq:V0}
  V_0 := \frac{\mathrm{Da}^2 R^3}{6\pi \eta \beta^2 \lambda^2} 
\end{equation}  
for the velocity scale,
\begin{equation}
\label{eq:Gr}
  \bG^{\perp}_{\ell m; \ell' m'} := 
  \int d\Omega\; \mathbf{e}_r (\theta,\varphi) \, 
  Y_{\ell m}(\theta,\varphi) \, Y_{\ell' m'}(\theta,\varphi)
\end{equation}
and
\begin{equation}
\label{eq:Gp}
\bG^{\parallel}_{\ell m; \ell' m'} :=  - \frac{1}{\ell'+1}
  \int d\Omega\;
  Y_{\ell m}(\theta,\varphi) \, \tilde{\nabla}_\parallel 
  Y_{\ell' m'}(\theta,\varphi) 
\end{equation}
for the geometrical factors (with $\tilde{\nabla}_\parallel$ defined by 
\eq{eq:nablapar1}),
and
\begin{equation}
  \label{eq:gr}
  g^{\perp}_{\ell \ell'} (\tilde{\xi}) 
  := \int_1^\infty ds\; 
  \left( - \frac{1}{2s^3} + \frac{3}{2s} - 1
  \right) s^{-\ell'} \, c_\ell(s,\tilde{\xi}) 
\end{equation}
and
\begin{equation}
  \label{eq:gp}
  g^{\parallel}_{\ell \ell'} (\tilde{\xi}) 
  := 
  \int_1^\infty ds\; 
  \left( \frac{1}{4 s^3} + \frac{3}{4 s} - 1  \right)
  s^{-\ell'} \, c_\ell(s,\tilde{\xi})
\end{equation}
for the dimensionless functions $g$ that encode the dependence on the
correlation length brought about by the solutions of the diffusion and
the hydrodynamical problems (see \eq{eq:ccoeff} for the definition of
the coefficients $c_\ell$). These integrals can be expressed in terms
of the exponential integral function (see Sec.~\ref{sec:examples} for
an example).  The superscripts $\parallel$ and $\perp$ pertain to the
contribution by the tangential and the radial component of the driving
force, respectively. A glimpse at \eq{eq:Vpsimu} allows one to
rationalize the velocity scale $V_0$, defined by \eq{eq:V0}, in terms
of the characteristic scale $\mathrm{Da}^2 R/(\beta \lambda)^2$ of the
source term $n_0 \Psi \nabla \mu$ (already discussed in connection
with Eqs.~(\ref{eq:deltamu}) and (\ref{eq:Psi})) and in terms of the
scale $R^2/(6\pi\eta)$ associated with the hydrodynamic kernel.

One can employ the same procedure in order to compute the angular velocity: if 
\eq{eq:deltan} is inserted into the expression for the angular velocity given 
by \eq{rot_self_propel}, after dropping the perfect gradient terms one 
obtains
\begin{eqnarray}
  \label{eq:Omegapsimu}
  \bOmega 
  & =
  & \frac{1}{8 \pi \eta R^3} \int\limits_{r>R} 
  \vol\;
    \left[ \left(\frac{R}{r}\right)^3 - 1 \right] \mathbf{r} \times
    \left[ - \delta n (\br) \, \nabla \mu(\br) \right]
    \nonumber
  \\
  & =
  & \frac{1}{8 \pi \eta R^3} \int\limits_{r>R} 
  \vol\;
    \left[ \left(\frac{R}{r}\right)^3 - 1 \right] \mathbf{r} \times
  \left[ - n_0 \Psi(\br) \, \nabla \mu(\br) \right] .
\end{eqnarray}
Inserting the decomposition~(\ref{eq:fradpar}) into this expression, one 
likewise arrives at 
\begin{equation}
  \label{eq:tildeomega}
  \bOmega = \Omega_0
  \sum_{\ell m} \sum_{\ell' m'} 
  a_{\ell m} a_{\ell' m'} \, 
  g^{\tau}_{\ell \ell'} \left(\frac{\xi}{R}\right)\,
  \bG^{\tau}_{\ell m; \ell' m'} ,
  \qquad
  \Omega_0 := \frac{3 V_0}{4R} ,
\end{equation}
with the geometrical factor
\begin{equation}
  \label{eq:Gom}
  \bG^{\tau}_{\ell m; \ell' m'} := - \frac{1}{\ell'+1}
  \int d\Omega\;
  Y_{\ell m}(\theta,\varphi) \, \be_r (\theta,\varphi) \times
  \tilde{\nabla}_\parallel Y_{\ell' m'}(\theta,\varphi)
\end{equation}
and the dimensionless function
\begin{equation}
  \label{eq:gom}
  g^{\tau}_{\ell \ell'} (\tilde{\xi}) :=
  \int_1^\infty ds\; 
  \left( \frac{1}{s^3} - 1 \right) s^{1-\ell'} \,
  c_\ell(s,\tilde{\xi}) ,
\end{equation}
where the superscript $\tau$ reminds that these factors arise from the driving 
torque $\br\times\mathbf{f}(\br)$. As will be shown in Sec.~\ref{sec:geometry} 
below, it turns out that $\bOmega \equiv 0$, regardless of the form of the 
activity pattern described by the amplitudes $a_{\ell m}$.

\section{The geometrical constraints}
\label{sec:geometry}

We define the spherical harmonics following the standard convention
employed in quantum mechanics \cite{messiah}: 
\begin{equation}
  \label{eq:Ylm}
  Y_{\ell m}(\theta,\varphi) = (-1)^m \sqrt{\frac{2\ell+1}{4\pi}
    \frac{(\ell-m)!}{(\ell+m)!}} P_{\ell}^m (\cos\theta) \mathrm{e}^{i
    m\varphi} ,
  \qquad
  m \geq 0,
\end{equation}
and 
\begin{equation}
  \label{eq:Ylm*}
  Y_{\ell, -m}(\theta,\varphi) = (-1)^m Y_{\ell m}^*(\theta,\varphi)
\end{equation}
in terms of the associated Legendre functions \cite{messiah}
\begin{equation}
  P_\ell^m(u) = (1-u^2)^{m/2} \frac{d^m}{du^m} P_\ell(u) 
  = \frac{1}{\ell! \, 2^\ell} (1-u^2)^{m/2}
  \frac{d^{\ell+m}}{du^{\ell+m}} (u^2-1)^\ell .
\end{equation}
We introduce the orbital angular momentum operator $\bL$, which
satisfies (upon setting the Planck constant $\hbar=1$)
\begin{equation}
  \label{eq:opL}
  \bL = \be_z L_z + \frac{\be_x-i\be_y}{2} L_+ 
  + \frac{\be_x+i\be_y}{2} L_- 
\end{equation}
with $L_z = -i \,\partial_\phi$ and $L_\pm = \mathrm{e}^{\pm i\,\phi} (\pm 
\partial_\theta + i \, \cot \theta \, \partial_\phi)$, so 
that 
\begin{equation}
  L_z Y_{\ell m} = m Y_{\ell m} 
\end{equation}
and
\begin{equation}
  L_{\pm} Y_{\ell m} = \sqrt{\ell(\ell+1)-m(m\pm 1)} 
  \; Y_{\ell m\pm 1} .
\end{equation}
In order to evaluate the integrals in Eqs.~(\ref{eq:Gr}) and (\ref{eq:Gp}), we 
make use of the following relations:
\begin{equation}
  \bL = r \be_r\times (-i \nabla) = -i \be_r\times\tilde{\nabla}_\parallel 
\end{equation}
and
\begin{equation}
  \label{eq:nablap}
  \tilde{\nabla}_\parallel = - i \be_r \times \bL ,
\end{equation}
with the operator $\tilde{\nabla}_\parallel$ given by \eq{eq:nablapar1}.  In 
order to evaluate \eq{eq:Gr}, we employ the following representation of the 
unit radial vector:
\begin{equation}
  \label{eq:er}
  \be_r = \sqrt{\frac{4\pi}{3}} \left[ \be_z \, Y_{1,0} 
    - \frac{\be_x-i\be_y}{\sqrt{2}} \, Y_{1,+1}
    + \frac{\be_x+i\be_y}{\sqrt{2}} \, Y_{1,-1}
    \right] ,
\end{equation}
and we use the identity \cite{messiah}
\begin{equation}
  \int d\Omega\; Y_{1M} \, Y_{\ell m} \, Y_{\ell' m'}
  = \sqrt{\frac{3 (2\ell+1)(2\ell'+1)}{4\pi}} 
  \left(
    \begin{array}[c]{ccc}
      1 & \ell & \ell' \\
      0 & 0 & 0 
    \end{array}
  \right)
  \;
    \left(
    \begin{array}[c]{ccc}
      1 & \ell & \ell' \\
      M & m & m' 
    \end{array}
  \right) ,
\end{equation}
in terms of the Wigner 3j symbols. It is then straightforward to obtain
\begin{eqnarray}
  \label{eq:Grevaluated}
  \bG^\perp_{\ell m; \ell' m'} 
  & = 
  & \sqrt{(2\ell+1)(2\ell'+1)} 
    \left(
    \begin{array}[c]{ccc}
      1 & \ell & \ell' \\
      0 & 0 & 0 
    \end{array}
  \right)
  \;
  \\
  & 
  &     
    \times \left[ \be_z \, \left(
    \begin{array}[c]{ccc}
      1 & \ell & \ell' \\
      0 & m & m' 
    \end{array}
   \right) 
   - \frac{\be_x-i\be_y}{\sqrt{2}} \,
    \left(
    \begin{array}[c]{ccc}
      1 & \ell & \ell' \\
      1 & m & m' 
    \end{array}
    \right) 
    + \frac{\be_x+i\be_y}{\sqrt{2}} \,
    \left(
    \begin{array}[c]{ccc}
      1 & \ell & \ell' \\
      -1 & m & m' 
    \end{array}
  \right) 
\right] .
 \nonumber 
\end{eqnarray}
The properties of the 3j symbols lead to the conclusion that 
$\bG^\perp_{\ell m; \ell' m'}$ can be non-zero only if $|\ell-\ell'|\leq 1$ 
(excluding the case $\ell=\ell'=0$) and if $|m+m'|\leq 1$. Furthermore, the 
3j symbol which has been factored  out vanishes if $\ell=\ell'$, and thus the 
case $|\ell-\ell'|=0$ actually does not arise.

Combining Eqs.~(\ref{eq:opL}, \ref{eq:nablap}, \ref{eq:er}), one finds
\begin{eqnarray}
  \tilde{\nabla}_\parallel 
  & = 
  & \sqrt{\frac{4\pi}{3}} \left[ 
    \frac{\be_x-i\be_y}{\sqrt{2}} Y_{1,1} 
    + \frac{\be_x+i\be_y}{\sqrt{2}} Y_{1,-1} 
    \right] L_z 
    \nonumber
  \\
  & 
  & \mbox{} + 
    \sqrt{\frac{2\pi}{3}} \left[ 
    \frac{\be_x-i\be_y}{\sqrt{2}} Y_{1,0} 
    - \be_z \, Y_{1,-1} 
    \right] L_+ 
    \nonumber
  \\
  & 
  & \mbox{} - 
    \sqrt{\frac{2\pi}{3}} \left[ 
    \frac{\be_x+i\be_y}{\sqrt{2}} Y_{1,0} 
    + \be_z \, Y_{1,+1} 
    \right] L_- . 
\end{eqnarray}
Inserting this expression into \eq{eq:Gp} and evaluating it as above, one 
finally arrives at 
\begin{eqnarray}
  \bG^\parallel_{\ell m; \ell' m'} 
  & = 
  & - \frac{\sqrt{(2\ell+1)(2\ell'+1)}}{\ell'+1}
    \left(
    \begin{array}[c]{ccc}
      1 & \ell & \ell' \\
      0 & 0 & 0 
    \end{array}
  \right)
  \\
  & 
  & \times \left\{ m' 
    \left[ 
    \frac{\be_x+i\be_y}{\sqrt{2}} 
    \left(
    \begin{array}[c]{ccc}
      1 & \ell & \ell' \\
      -1 & m & m' 
    \end{array}
  \right)
   + \frac{\be_x-i\be_y}{\sqrt{2}} 
    \left(
    \begin{array}[c]{ccc}
      1 & \ell & \ell' \\
      1 & m & m' 
    \end{array}
  \right)
  \right]
  \right.
 \nonumber
 \\
&
& + \frac{\sqrt{\ell'(\ell'+1)-m'(m'+1)}}{\sqrt{2}}
  \left[ 
  \frac{\be_x-i\be_y}{\sqrt{2}}
  \left(
    \begin{array}[c]{ccc}
      1 & \ell & \ell' \\
      0 & m & m'+1 
    \end{array}
  \right)
 - \be_z \, \left(
    \begin{array}[c]{ccc}
      1 & \ell & \ell' \\
      -1 & m & m'+1 
    \end{array}
  \right)
 \right]
 \nonumber
  \\
&
& \left. + \frac{\sqrt{\ell'(\ell'+1)-m'(m'-1)}}{\sqrt{2}}
  \left[ 
    -\frac{\be_x+i\be_y}{\sqrt{2}}
  \left(
    \begin{array}[c]{ccc}
      1 & \ell & \ell' \\
      0 & m & m'-1 
    \end{array}
  \right)
 - \be_z \, \left(
    \begin{array}[c]{ccc}
      1 & \ell & \ell' \\
      1 & m & m'-1 
    \end{array}
  \right)
 \right]
 \right\} .
 \nonumber
 \end{eqnarray}
The same conditions for the indices as in \eq{eq:Grevaluated} also hold for 
this expression.

Concerning the geometrical factors entering into the expression of the angular 
velocity, \eq{eq:Gom} can be written as
\begin{eqnarray}
  \label{eq:Gom2}
  \bG^{\tau}_{\ell m; \ell'm'} 
  & = 
  & -\frac{i (-1)^m}{\ell'+1}
    \int d\Omega\; 
    Y_{\ell, -m}^* \, \bL Y_{\ell'
    m'}
    \nonumber
  \\
  & = 
  & -\frac{i (-1)^m}{\ell'+1} \delta_{\ell,\ell'}
    \left[ \be_z \, m' \delta_{m+m',0} \right.
    \nonumber
  \\
  & 
  & \left. \mbox{} + \sqrt{\ell(\ell+1)+m m'}
    \left( 
    \frac{\be_x-i\be_y}{2} \, \delta_{m+m',-1} 
    + \frac{\be_x+i\be_y}{2} \, \delta_{m+m',1} 
    \right)
    \right] .
\end{eqnarray}
This result implies that the angular velocity given by \eq{eq:tildeomega} 
vanishes identically: the activity associated with a given multipole 
$\ell$ is 
\begin{equation}
  \label{eq:Aell}
  \mathbb{A}_\ell(\theta,\varphi) := \sum_{m=-\ell}^\ell a_{\ell m}
  Y_{\ell m}(\theta,\varphi), 
\end{equation}
while the constraint $\delta_{\ell,\ell'}$ in \eq{eq:Gom2} implies
that \eq{eq:tildeomega} can be written as
\begin{equation}
  \label{eq:tildeomega2}
  \bOmega = \Omega_0
  \sum_{\ell} \frac{g^{\tau}_{\ell\ell}
    (\tilde{\xi})}{\ell +1}
  \bomega_\ell , 
\end{equation}
with the vector
\begin{equation}
  \label{eq:diagonalElement}
  \bomega_\ell :=
  \int d\Omega\; \mathbb{A}_\ell^* \, (i \bL) \,
  \mathbb{A}_\ell \, ,
\end{equation}
after taking into account the identity
$a_{\ell, m}^* = (-1)^m a_{\ell, -m}$ for the real function
$\mathbb{A}(\theta,\varphi)$.  Written in this form, it is clear that
each component of the vector $\bomega_\ell$ is actually the diagonal
matrix element of the corresponding component of the anti-Hermitian
operator $i\bL$ (i.e., the generator of rotations), and therefore it
must be imaginary.  But the angular velocity $\bOmega$ as given by
\eq{eq:tildeomega2} must be real. Consequently, the only possibility
is that it is zero. We emphasize that the symmetric dependence on the
activity--induced inhomogeneity is crucial for the conclusion, as
illustrated by Eqs.~(\ref{eq:curlf}) or (\ref{eq:symetricOmega}). This
differs from the standard model of phoresis, in which the expression
equivalent to \eq{eq:diagonalElement} is an off-diagonal matrix
element, involving on one hand the angular dependence of the activity
pattern, and on the other hand the angular dependence of the
interaction potential.

\section{The limit of a short correlation length ($\xi \ll R$)}
\label{sec:Lambda0}

In order to evaluate the transport velocity in this limit, it is
physically more insightful and mathematically more straightforward to
start with the general expression in \eq{eq:Vpsimu} rather than with
the expansion in \eq{eq:driftV}. In the limit $\xi/R \ll 1$, one
infers from the explicit solution in \eq{eq:Psi} that the field
$\Psi(\br)$ is exponentially confined to the radial region of
extension $\sim \xi$ around the particle (because the Bessel functions
appearing in the expansion decay exponentially for large values of
their argument \cite{abram,GrRy94}), while the chemical potential in
\eq{eq:deltamu} is insensitive to this limit. Therefore, the integral
in \eq{eq:Vpsimu} is exponentially dominated by the behavior of the
integrand near the lower limit $r=R$. One can apply Watson's lemma
\cite{BeOr78} in order to estimate the integral by approximating the
whole non-exponential dependence of the integrand by the expansion
around $r=R$.  For this purpose, we use \eq{eq:deltan} in
\eq{eq:Vpsimu} and drop an irrelevant gradient term, so that
\begin{equation}
  \bV =
  - \frac{n_0}{6\pi\eta R} \left(\frac{\lambda}{\xi}\right)^2 
  \int_{r>R} 
  \vol\;
  \Psi(\br) 
  \left[ \Ttens(\br) - \Itens \right] 
  \cdot\nabla n(\br) .
\end{equation}
The hydrodynamic tensor in the integrand is approximated by a Taylor
expansion as
\begin{equation}
  \label{eq:Tsmallr}
  \Ttens(\br) - \Itens \approx
  (r-R)\be_r\cdot\nabla\Ttens(R\be_r)
  = \frac{3}{2} \left(1-\frac{r}{R}\right) 
  \left( \Itens - \be_r\be_r \right),
\end{equation}
because the difference $\Ttens(\br) - \Itens$ vanishes at
$r=R$ due to the no-slip boundary condition~(\ref{eq:noslip}). This
result implies the approximation
\begin{equation}
  \left[ \Ttens(\br) - \Itens \right] 
  \cdot\nabla n(\br) 
  \approx
  \frac{3}{2} \left(1-\frac{r}{R}\right) \nabla_{\parallel} n(R\be_r) ,
\end{equation}
where we have introduced the operator 
\begin{equation}
  \nabla_{\parallel} := \left. \left( \Itens - \be_r\be_r
    \right)\cdot \nabla \right|_{r=R} 
  = \frac{1}{R} \tilde{\nabla}_\parallel 
\end{equation}
(in terms of the operator introduced in \eq{eq:nablapar1}), which represents 
the tangential gradient on the sphere of radius $R$. Therefore, the transport 
velocity can be approximated as
\begin{equation}
  \label{eq:thinV}
  \bV
  \approx 
  - \frac{n_0}{6\pi\eta R} \left(\frac{\lambda}{\xi}\right)^2 
  \int d\Omega \;
  \int_R^\infty dr\; R^2 \, \Psi(\br) \, 
  \frac{3}{2} \left(1-\frac{r}{R}\right) \nabla_{\parallel}
  n(R\be_r)
  = - \frac{1}{4\pi} \int d\Omega \; \bv_\mathrm{slip}(\theta,\varphi)
\end{equation}
with (see \eq{eq:vslip} in the main text)
\begin{equation}
  \label{eq:vslipSI}
  \bv_\mathrm{slip} (\theta,\varphi) := \mathcal{L}(\theta,\varphi) \,
  \nabla_{\parallel}  n(R\be_r)
\end{equation}
and
\begin{equation}
  \label{eq:Lgeneral}
  \mathcal{L}(\theta,\varphi) :=
  \frac{n_0}{\eta} \left(\frac{\lambda}{\xi}\right)^2
  \int_R^\infty dr\; (R-r) \Psi(r,\theta,\varphi) .
\end{equation}
An interpretation of this result is that the hydrodynamical effect of
the activity can be modeled as a local boundary condition at each
point of the surface of the particle in terms of a slip velocity given
by $\bv_\mathrm{slip}$ with a phoretic coefficient
$\mathcal{L}(\theta,\varphi)$.  One can now evaluate the expression
for the phoretic coefficient by inserting the explicit solution (see
Eqs.~(\ref{eq:Psi}--\ref{eq:psiell})) and taking the limit
$\tilde{\xi} := \xi/R \to 0$:
\begin{eqnarray}
  \mathcal{L}(\theta,\varphi) 
  & =
  & - \mathrm{Da}\, \frac{R^4}{\eta \beta\xi^2}
    \sum_{\ell m} a_{\ell m} \, Y_{\ell m}(\theta,\varphi) 
    \int_1^\infty ds\; \frac{1-s}{\sqrt{s}} 
    \frac{\tilde{\xi}^2\, K_{\ell+1/2}(s \, \tilde{\xi}^{-1})}{(\ell+1) K_{\ell+1/2}(\tilde{\xi}^{-1})
    + \tilde{\xi}^{-1} K_{\ell-1/2}(\tilde{\xi}^{-1})} 
    \nonumber 
  \\
  & \stackrel{\tilde{\xi}\to 0}{\approx}
  & - \mathrm{Da}\, \frac{R^2 \tilde{\xi}}{\eta \beta}
    \sum_{\ell m} a_{\ell m} \, Y_{\ell m}(\theta,\varphi) 
    \int_1^\infty ds\; \frac{1-s}{s} \mathrm{e}^{(1-s)/\tilde{\xi}}
    \nonumber 
  \\
  & \stackrel{\tilde{\xi}\to 0}{\approx}
  & - \mathrm{Da}\, \frac{R^2 \tilde{\xi}}{\eta \beta}
    \sum_{\ell m} a_{\ell m} \, Y_{\ell m}(\theta,\varphi) 
    \int_1^\infty ds\; (1-s) \mathrm{e}^{(1-s)/\tilde{\xi}}
    \nonumber 
  \\
  & = 
  & \mathrm{Da}\, \frac{R^2 \tilde{\xi}^3}{\eta \beta}
    \sum_{\ell m} a_{\ell m} \, Y_{\ell m}(\theta,\varphi) ,
\end{eqnarray}
where for $\tilde{\xi} \to 0$ the Bessel functions have been expanded
uniformly inside the integral, and Watson's lemma \cite{BeOr78} was
used to approximate the non-exponential contributions in the integrand
around $s=1$.  If the last line is evaluated by using the definition
for the activity function (\eq{eq:alm} in the main text), one arrives
at the final result for the phoretic coefficient (see \eq{eq:L} in the
main text),
\begin{equation}
  \label{eq:LSI}
  \mathcal{L}(\theta,\varphi) :=
  \frac{\mathrm{Da} \, \xi^3}{\eta\beta R} \,
  \mathbb{A}(\theta,\varphi) .
\end{equation}

In order to complete the analysis, we address the limit $\xi\to 0$ of
the field $\nabla_\parallel n$. It is clear from \eq{eq:Vpsimu} that
one could have derived a similar expression for the slip velocity
given by \eq{eq:vslipSI} but with $\nabla_\parallel n$ replaced by
$(\xi/\lambda)^2 \nabla_\parallel \mu$. Since the distribution of the
chemical potential induced by the activity is actually independent of
the correlation length (see \eq{eq:deltamu}), the phoretic velocity
will scale as $V\sim\xi^5$ in the limit $\xi/R\to 0$ (see
\eq{eq:LSI}).

Although the angular velocity $\bOmega$ has been shown to vanish identically, 
it is instructive and useful for future reference to compute also the 
approximation for $\bOmega$ in this limit. The procedure is identical: we use 
\eq{eq:deltan} in \eq{eq:Omegapsimu} and drop an irrelevant gradient term, so 
as to obtain
\begin{equation}
  \bOmega
  = - \frac{n_0}{8\pi\eta R^3} \left(\frac{\lambda}{\xi}\right)^2 
  \int\limits_{r>R} 
  \vol\;
  \Psi(\br)
  \left[ \left(\frac{R}{r}\right)^3 - 1 \right] \mathbf{r} \times \nabla n(\br) .
\end{equation}
Approximating the integrand as before, one arrives at
\begin{equation}
  \label{eq:thinOmega}
  \bOmega
  = - \frac{1}{4 \pi} \int d\Omega\;
  \bomega_\mathrm{slip}(\theta,\varphi) ,
\end{equation}
with a ``slip angular velocity'' 
\begin{equation}
  \label{eq:omegaslip}
  \bomega_\mathrm{slip}(\theta,\varphi)
  = \frac{3}{2 R}
  \mathcal{L} (\theta,\varphi)\, \be_r\times \nabla_\parallel n(R\be_r) .  
\end{equation}
Although $\mathcal{L}\neq 0$ in general, and the relationship between
$\bomega_\mathrm{slip}$ and $\bv_\mathrm{slip}$ (see \eq{eq:vslip}) agrees with 
the standard result for phoresis \cite{Anderson1989}, the angular integral in 
\eq{eq:thinOmega} over $\bomega_\mathrm{slip}$ is identically zero because it 
was demonstrated in Sec.~\ref{sec:geometry} that $\bOmega \equiv 0$.

\section{The limit of a large correlation length ($\xi \gg R$)}
\label{sec:Lambdainfty}

The behavior of the transport velocity in the limit
$\tilde{\xi} := \xi/R\gg 1$ can be addressed by considering the
behavior of the radial integrals (see Eqs.~(\ref{eq:gr}) and
(\ref{eq:gp})). It turns out that they exhibit a finite limit as
$\tilde{\xi}\to\infty$, except for the terms representing the
monopole--dipole coupling in \eq{eq:driftV}: they require a separate
analysis because the diffusion problem is ill-posed for these modes in
the limit $\tilde{\xi}\to\infty$.

Instead of dealing with the expressions in Eqs.~(\ref{eq:gr}) and
(\ref{eq:gp}), it is actually more useful to consider the expression
analogous to \eq{eq:driftV}, but starting from \eq{eq:V} rather than
from \eq{eq:Vpsimu}: although both expressions agree, the alternative
one is mathematically less involved upon considering the limit
$\tilde{\xi}\to\infty$. If this alternative is adopted, one obtains an
expansion similar to \eq{eq:driftV}, but with new radial integrals:
\begin{equation}
  \label{eq:hatgr}
  \hat{g}^{\perp}_{\ell \ell'} (\tilde{\xi}) 
  := \int_1^\infty ds\; 
  \left( - \frac{1}{2s^3} + \frac{3}{2s} - 1
  \right) s^{-\ell'} \, \hat{c}_\ell(s, \tilde{\xi}) 
\end{equation}
and
\begin{equation}
  \label{eq:hatgp}
  \hat{g}^{\parallel}_{\ell \ell'} (\tilde{\xi}) 
  := \int_1^\infty ds\; 
  \left( \frac{1}{4 s^3} + \frac{3}{4 s} - 1  \right)
  s^{-\ell'} \, \hat{c}_\ell(s,\tilde{\xi}) ,
\end{equation}
with the coefficients $\hat{c}_\ell$ stemming from the expansion for
the field $\delta n(\br)$ (see \eq{eq:hatccoeff}) rather than from the
field $\Psi(\br)$ (see \eq{eq:ccoeff}). One has to consider only those
integrals for which the pair $(\ell,\ell')$ satisfies the geometrical
constraint $|\ell-\ell'|=1$ (because these are the only terms
contributing to the sum in \eq{eq:driftV}, see
Sec.~\ref{sec:geometry}). In this case the new integrals do converge at
the upper limit of integration in spite of the algebraic decay of the
new factor $\hat{c}_\ell(s,\tilde{\xi})$ as $s\to\infty$, with the
worst case corresponding to the pairs $(\ell,\ell')=(0,1)$ or $(1,0)$,
which are associated with the monopole--dipole coupling in the sum in
Eq.~(\ref{eq:driftV}).

It can be proven that these integrals converge \emph{uniformly} as
functions of $\tilde{\xi}$ (see the subsection below), so that
the limit $\tilde\xi\to\infty$ can be taken inside the integrals. In
this limit one has
\begin{equation}
  \label{eq:hatd}
  \hat{c}_\ell(s,\tilde{\xi})
  = \hat{d}_\ell(s)
  + \left\{
    \begin{array}[c]{cl}
      \displaystyle \tilde{\xi} + o(\tilde{\xi}^{-1}) , & \ell = 0, \\
      \\
      \displaystyle o(\tilde{\xi}^{-1}) , & \ell = 1, \\
      \\
      \displaystyle o(\tilde{\xi}^{-2}) , & \ell \geq 2, \\
    \end{array}
  \right.
  \qquad
  \hat{d}_\ell(s) := \frac{s^{1-\ell}}{2 (2\ell-1) (\ell+1)}\left( 1 -
    \frac{\ell-1}{\ell+1} s^{-2} \right) .
\end{equation}
Thus one obtains
\begin{eqnarray}
  \label{eq:hatgrlimit}
  \lim_{\tilde{\xi}\to\infty} g^\perp_{\ell\ell'}(\tilde{\xi})
  & =
  & 
    \int_1^\infty ds\;
  \left( - \frac{1}{2s^3} + \frac{3}{2s} - 1
    \right)
    s^{-\ell'} \, \hat{d}_\ell(s) 
    \nonumber
  \\
  & = &     
  \frac{3(\ell-1)-3(\ell+\ell')(4\ell+\ell')}{
(2\ell-1)(\ell+1)^2(\ell+\ell')(\ell+\ell'-1)(\ell+\ell'+1)(\ell+\ell'-2)(\ell+
\ell'+3)} ,
\end{eqnarray}
and
\begin{eqnarray}
  \label{eq:hatgplimit}
  \lim_{\tilde{\xi}\to\infty} g^\parallel_{\ell\ell'}(\tilde{\xi})
  & =
  & 
    \int_1^\infty ds\;
  \left( \frac{1}{4s^3} + \frac{3}{4s} - 1
    \right)
    s^{-\ell'}\, \hat{d}_\ell(s) 
    \nonumber
  \\
  & = &       
  -\frac{6(\ell+1)+3(3\ell+\ell')(\ell+\ell'-1)(\ell+\ell'+2)}{
2(2\ell-1)(\ell+1)^2(\ell+\ell')(\ell+\ell'-1)(\ell+\ell'+1)(\ell+\ell'-2)(\ell+
\ell'+3)} 
\end{eqnarray}
for all pairs $(\ell,\ell')$ that satisfy $|\ell-\ell'|=1$, except for
the pairs $(0,1)$ and $(1,0)$ --- the latter does occur because the
asymptotic behaviors of $\hat{d}_0(s)$ and $\hat{d}_1(s)$,
respectively, do not decay with increasing $s$ and the corresponding
integrals diverge. The resulting expressions~(\ref{eq:hatgrlimit})
and (\ref{eq:hatgplimit}) decay so quickly upon increasing $\ell$,
that they yield a convergent contribution to the sum in
Eq.~(\ref{eq:driftV}) which determines the phoretic velocity of the
particle.

At this stage, one can understand the rationale behind the
redefinitions in Eqs.~(\ref{eq:hatgr}) and (\ref{eq:hatgp}): the term
that accounts for the difference between $c_\ell$ and $\hat{c}_\ell$
(see \eq{eq:hatccoeff}), and which is related to the chemical
potential, eliminates the dominant divergent contributions (of the
order of $\tilde{\xi^2}$) to the radial integrals. Those terms,
however, do not play any role because they correspond to a purely
potential force in the Stokes equation and cancel each other if the
sum in Eq.~(\ref{eq:driftV}) is carried out.

This kind of global cancellation is actually what occurs for the
monopole--dipole coupling, the analysis of which has been pending: one
can explicitly carry out the integrals in Eqs.~(\ref{eq:hatgr}) and
(\ref{eq:hatgp}) for the specific choices $(\ell,\ell')=(0,1)$ and
$(1,0)$ and one finds that, in spite of the redefinition, they still
exhibit diverging terms proportional to $\tilde{\xi}$ or
$\ln \tilde{\xi}$.  These divergences, however, drop out in the
combination
\begin{displaymath}
  \sum_m a_{00} a_{1m} \left[ 
    g^{\perp}_{10}(\tilde{\xi}) \bG^{\perp}_{1 m; 00} 
    + g^{\parallel}_{10}(\tilde{\xi}) \bG^{\parallel}_{1m; 00}
    + g^{\perp}_{01}(\tilde{\xi}) \bG^{\perp}_{00;1 m} 
    + g^{\parallel}_{01}(\tilde{\xi}) \bG^{\parallel}_{00;1m}
  \right] ,
\end{displaymath}
which contributes to the sum in Eq.~(\ref{eq:driftV}) and is finite
and nonzero in the limit $\tilde{\xi}\to\infty$, as illustrated
explicitly by the example worked out in Sec.~\ref{sec:examples}. In
order to understand this particularity of the monopolar and the
dipolar terms, one can consider the diffusion problem posed by
Eqs.~(\ref{eq:linmu}), (\ref{eq:bcnlin}), and (\ref{eq:deltamu}) for
the field $\delta n(\br)$: the coefficients $\hat{c}_\ell(r/R,\xi/R)$
given by \eq{eq:hatccoeff} satisfy the boundary value problem
\begin{equation}
  \frac{d^2 \hat{c}_\ell}{d r^2} + \frac{2}{r}
  \frac{d\hat{c}_\ell}{d r}- \frac{\ell(\ell+1)}{r^2} \hat{c}_\ell
  - \frac{\hat{c}_\ell}{\xi^2} = - \frac{1}{\ell+1}
  \left(\frac{R}{r}\right)^{\ell+1} ,
  \qquad
  \frac{d\hat{c}_\ell}{d r}(r=R) = 0 ,
  \qquad
  \lim_{r\to\infty} \hat{c}_\ell = 0 .
\end{equation}
For a coefficient with $\ell\geq 2$, in the limit $\xi\to\infty$ one
can drop the term $\hat{c}_\ell/\xi^2$ from the equation, and still
obtain a well--posed problem, the solution of which is precisely the
function $\hat{d}_\ell(r/R)$ (\eq{eq:hatd}). This is not the case for
the monopole ($\ell=0$) and the dipole ($\ell=1$), because the
inhomogeneity of the differential equation, stemming from the spatial
distribution (\eq{eq:deltamu}) of the chemical potential induced by
the activity, decays too slowly with separation, so that the boundary
condition at infinity cannot be fulfilled. Therefore, the monopolar
and dipolar modes of the density field require a large--distance
cutoff $\xi$ in order to be well defined. Ultimately, however, the
nontrivial conclusion is that this cutoff is nevertheless irrelevant
for the phoretic velocity of the particle, which attains a finite and
nonzero value in the limit $\xi\to\infty$, as has been shown above.

\subsubsection*{Uniform convergence of the integrals in
  Eqs.~(\ref{eq:hatgr}) and (\ref{eq:hatgp})}

In order to study the limit $\tilde\xi\to\infty$ of the integrals, we
prove uniform convergence. To this end, we derive bounds on the
coefficients $\hat{c}_\ell$:
one can rearrange the coefficients according to (see
Eqs.~(\ref{eq:ccoeff}) and (\ref{eq:hatccoeff}))
\begin{equation}
  \hat{c}_\ell(s,\tilde{\xi}) = \frac{\tilde{\xi}^2}{(\ell+1)
    s^{\ell+1}} \left[ 1 - \chi_\ell(\tilde{\xi}) \,
    \kappa_\ell\left(\frac{s}{\tilde{\xi}}\right)
  \right] ,
\end{equation}
with
$\chi_\ell (\tilde{\xi}) := - (\ell+1) \tilde{\xi}^{\ell-3/2} \,
\psi_\ell(\tilde{\xi}) \geq 0$ (see \eq{eq:psiell})
and
\begin{equation}
  \kappa_\ell(z) := z^{\ell+1/2} \, K_{\ell+1/2}(z) ,
  \qquad
  \kappa_\ell'(z) = - z^{\ell+1/2} \, K_{\ell-1/2}(z) .
\end{equation}
(The last result follows from the recursion relation
$\left[ z^p K_p(z) \right]' = - z^p K_{p-1}(z)$ \cite{abram,GrRy94}.)
Since $K_p(z)$ is a positive function that approaches zero
exponentially, one concludes that $\kappa_\ell(z)$ is a positive,
monotonically decreasing function that approaches zero exponentially;
furthermore, we find
\begin{equation}
  \chi_\ell(\tilde{\xi}) \,
  \kappa_\ell(\tilde{\xi}^{-1})
  = \frac{(\ell+1) K_{\ell+1/2}(\tilde{\xi}^{-1})}{(\ell+1)
    K_{\ell+1/2}(\tilde{\xi}^{-1})
    + \tilde{\xi}^{-1} K_{\ell-1/2}(\tilde{\xi}^{-1})}
  \leq 1 .
\end{equation}
Consequently, one has
\begin{equation}
  0 \leq \frac{\tilde{\xi}^2}{(\ell+1)
    s^{\ell+1}} \left[ 1 - \chi_\ell(\tilde{\xi}) \,
    \kappa_\ell(\tilde{\xi}^{-1})
  \right]
  \leq \hat{c}_\ell(s,\tilde{\xi})
  \leq \frac{\tilde{\xi}^2}{(\ell+1)
    s^{\ell+1}} 
\end{equation}
within the relevant integration range $1 \leq s < \infty$;
accordingly, the radial integrals in Eqs.~(\ref{eq:hatgr}) and
(\ref{eq:hatgp}) are bounded as
\begin{equation}
  \tilde{\xi}^{-2} \left| \hat{g}^{\perp}_{\ell \ell'}
    (\tilde{\xi}) \right|
  \leq 
  \frac{1}{(\ell+1)} \int_1^\infty ds\; 
  \left| - \frac{1}{2s^3} + \frac{3}{2s} - 1
  \right| s^{-\ell'-\ell-1} 
\end{equation}
and
\begin{equation}
  \tilde{\xi}^{-2} \left| \hat{g}^{\parallel}_{\ell \ell'}
    (\tilde{\xi}) \right|
  \leq 
  \frac{1}{(\ell+1)} \int_1^\infty ds\; 
  \left| \frac{1}{4s^3} + \frac{3}{4s} - 1
  \right| s^{-\ell'-\ell-1} .
\end{equation}
Both bounding integrals converge for any pair $(\ell,\ell')$ which
satisfies $|\ell-\ell'|=1$. Therefore, the integrals defining the
functions $\tilde{\xi}^{-2} g^\perp_{\ell\ell'}(\tilde{\xi})$ and
$\tilde{\xi}^{-2} g^\parallel_{\ell\ell'}(\tilde{\xi})$ converge
\textit{uniformly} as functions of $\tilde{\xi}$.  This allows one to
take the limit $\tilde{\xi}\to\infty$ inside the integrals for the
functions $g^\perp_{\ell\ell'}(\tilde{\xi})$ and
$g^\parallel_{\ell\ell'}(\tilde{\xi})$. The limit of these integrals
either exists or diverges not faster than $\tilde{\xi}^2$.

\section{Phoresis with $\lambda=0$ and $W\neq 0$}
\label{sec:W}

In the absence of correlations (i.e., $\lambda=0$) but with a spatially
extended interaction between the particle and the molecules of the 
chemical (i.e., $W \neq 0$), the chemical potential is given by \eq{eq:mu} as
\begin{equation}
  \label{eq:mu_nolambda}
  \mu(\br) = h'(n(\br)) + W(\br) .
\end{equation}
The phoretic velocity is still given by \eq{eq:V}, but now the deviation 
$\delta n = n-n_0$ from homogeneity is determined \emph{both} by the
activity $\mathbb{A}$ and the spatially extended interaction $W$. (We note 
that a non-vanishing drift still requires a non-zero activity, so that 
$\nabla\mu\neq 0$ in \eq{eq:V}).

We follow the same strategy as in the case $\lambda\neq 0$ and $W=0$: 
first the diffusion problem is solved by linearization in terms of 
the deviations $\delta n$, and then one evaluates \eq{eq:V} with the 
solutions which have been found. In this case, \eq{eq:linmu} is replaced 
by the linearized form of \eq{eq:mu_nolambda}, i.e.,
\begin{equation}
  \label{eq:linmuW}
  \mu \approx \mu_0 + h''(n_0)
  \, \delta n +W ,
\end{equation}
assuming that $W$ vanishes at infinity.  The chemical potential is
obtained as the solution for the same boundary--value problem
(\eqs{eq:laplacianmu} and (\ref{eq:bcmulin})) as before, i.e.,
\eq{eq:deltamu} holds, and the density field follows from
\eq{eq:linmuW}:
\begin{equation}
  \label{eq:deltanW}
  \delta n \approx \frac{1}{h''(n_0)} 
  \left( \mu -\mu_0 - W \right). 
\end{equation}
(We note that the boundary conditions in \eq{eq:bcnlin} are irrelevant
if $\lambda=0$.) Inserting this expression into \eq{eq:V}, and noting
that the term involving $\mu-\mu_0$ becomes a perfect gradient that
does not contribute, one finally arrives at
\begin{equation}
  \label{eq:VweakW}
  \bV^{(W)} =
  \frac{1}{6\pi\eta R} \int_{r>R} 
  \vol\;
  \left[ \Ttens(\br) - \Itens\right] \cdot 
  \left[ \frac{1}{h''(n_0)}
    W(\br) \, \nabla \mu(\br) \right] .
\end{equation}
(The superscript $(W)$ is a reminder that the expression corresponds to the 
case $W \neq 0$ and $\lambda = 0$.) This expression has precisely the same 
form as the one in \eq{eq:Vpsimu} with the identification
\begin{equation}
  \label{eq:psiW}
  n_0 \Psi(\br) \leftrightarrow
  - \frac{1}{h''(n_0)}
  W(\br).
\end{equation}
Therefore, by expanding the interaction field $W(\br)$ in terms of the 
spherical harmonics,
  \begin{equation}
    \label{eq:wlm}
    W(\br) = \sum_{\ell=0}^{\infty} \sum_{m=-\ell}^\ell 
    w_{\ell m}(r) Y_{\ell m}(\theta, \varphi) ,
  \end{equation}
  and by comparison with the expansion in \eq{eq:Psi}, one can infer
  the mapping
  \begin{equation}
    \label{eq:psitoW}
    \frac{\mathrm{Da}\, R^2}{\beta \lambda^2}
    c_\ell \left(\frac{r}{R}, \frac{\xi}{R} \right) a_{\ell m} 
    \quad \leftrightarrow \quad 
    - \frac{1}{h''(n_0)}
    w_{\ell m}(r)
\end{equation}
and borrow the final expression in
Eqs.~(\ref{eq:driftV}--\ref{eq:gp}). In this manner, one arrives
at the expression
\begin{equation}
  \label{eq:driftW}
  \bV^{(W)} = V_0^{(W)}
  \sum_{\ell m} \sum_{\ell' m'} 
  a_{\ell' m'} 
  \left[ 
    \gamma^{\perp}_{\ell m; \ell'} 
    \bG^{\perp}_{\ell m; \ell' m'} 
    +
    \gamma^{\parallel}_{\ell m; \ell'} 
    \bG^{\parallel}_{\ell m; \ell' m'}
  \right]
\end{equation}
with the velocity scale 
\begin{equation}
  \label{eq:V0w}
  V_0^\mathrm{(W)} =
  \frac{\mathrm{Da}\, R}{6\pi\eta\beta^2 h''(n_0)} 
\end{equation}
and the dimensionless radial integrals
\begin{equation}
  \label{eq:gammar}
  \gamma^{\perp}_{\ell m; \ell'} := - \int_1^\infty ds\;
  \left( -\frac{1}{2s^3} + \frac{3}{2s} - 1
  \right) s^{-\ell'} \, \beta w_{\ell m}\left( R s\right) ,
\end{equation}
and
\begin{equation}
  \label{eq:gammap}
  \gamma^{\parallel}_{\ell m; \ell'} := - \int_1^\infty ds\; 
  \left( \frac{1}{4 s^3} + \frac{3}{4 s} - 1  \right)
  s^{-\ell'} \, \beta w_{\ell m} \left( R s\right) .
\end{equation}
The velocity scale (\eq{eq:V0w}) is related to the corresponding one
in the correlation--induced model as
$V_0 = V_0^\mathrm{(W)} (R/\xi)^2 \mathrm{Da}$ (compare with
\eqs{eq:V0} and (\ref{eq:xi})). The dimensionless functions
$\gamma_{\ell m; \ell'}$ encode the ``shape'' of the interaction
$W(\br)$, whereby any length scale that characterizes the spatial
extent of $W$ would formally play the role of a cutoff and replace
$\xi$ in the mathematical expressions. The geometrical constraints
derived from the functions $\bG^\perp_{\ell m; \ell' m'}$,
$\bG^\parallel_{\ell m; \ell' m'}$ also affect the conclusions
following from \eq{eq:driftW}. In particular, a spherically symmetric
external potential $W(r)$ gives rise only to a monopolar contribution,
i.e., $w_{\ell m}=0$ if $\ell\neq 0$ in \eq{eq:wlm}. Therefore any
activity pattern lacking a dipole, i.e., $a_{\ell m}=0$ if $\ell=1$ in
the expansion in \eq{eq:alm}, cannot lead to phoresis: $\bV^{(W)}=0$
according to \eq{eq:driftW}. Likewise, in the complementary case of a
purely dipolar activity pattern ($a_{\ell m}=0$ if $\ell\neq 1$), the
correlation--induced velocity vanishes, but $\bV^{(W)}\neq 0$
generically for a spherically symmetric potential (it is
$\bG^\perp_{00;10}=-\bG^\parallel_{00;10}=\be_z/\sqrt{3}$, while
$\gamma^\perp_{00;\ell'}-\gamma^\parallel_{00;\ell'}\neq 0$ in
general.)

The same approach can be applied to obtain the phoretic angular
velocity: inserting \eq{eq:deltanW} into \eq{eq:Omegapsimu} allows one
to make again use of the identification in \eq{eq:psitoW}, so that
starting from the solution given in
\eqs{eq:tildeomega}--(\ref{eq:gom}) one can write
\begin{equation}
  \bOmega^{(W)} =
  \Omega_0^{(W)} \sum_{\ell m} \sum_{\ell' m'} 
  a_{\ell' m'} \, 
  \gamma^{\tau}_{\ell m; \ell'} 
  \bG^{\tau}_{\ell m; \ell' m'} ,
  \qquad \Omega_0^{(W)}:= \frac{3 V_0^{(W)}}{4 R},
\end{equation}
with the same geometrical factor as in \eq{eq:Gom} and with the
dimensionless radial integral
\begin{equation}
  \gamma^{\tau}_{\ell m; \ell'} = 
  - \int_1^\infty ds\;
  \left( \frac{1}{s^3} - 1 \right)
  s^{1-\ell'}
  \, \beta w_{\ell m}\left(R s\right) .
\end{equation}
(As above, the superscript $(W)$ is a reminder that the expression
corresponds to the case $W \neq 0$ and $\lambda = 0$.) In this case,
however, in general the angular velocity is non-zero. This is the case
because the argument presented in Sec.~\ref{sec:geometry}, which
relies fundamentally on the symmetric bilinear dependence on the
activity (see \eq{eq:diagonalElement}), is not applicable here.
    

Finally, for comparison with the case of a vanishing correlation
length addressed in Sec.~\ref{sec:Lambda0}, we consider the limit in
which the spatial extension of the potential $W(\br)$ vanishes, so
that the interaction between the particle and the molecules of the
chemical is localized at the surface $r=R$. The identification in
\eq{eq:psiW} still holds; if $W(\br)$ becomes exponentially small for
$r\neq R$, Watson's lemma can be applied again, so that the
expressions in \eqs{eq:thinV} and (\ref{eq:thinOmega}) in terms of the
slip velocities, \eq{eq:vslip} and \eq{eq:omegaslip}, are valid. The
phoretic coefficient is given by
\begin{equation}
  \label{eq:Lw}
  \mathcal{L}^{(W)} (\theta,\varphi) := \frac{1}{\eta}
  \int_R^\infty dr\; \left( r - R \right) 
  W (r,\theta,\varphi) ,
\end{equation}
as derived from \eq{eq:Lgeneral}. This coefficient agrees with the
limit of small $W$ for the expression derived in the case of a very
dilute chemical (so that the ideal gas approximation holds, see, e.g.,
Ref.~\cite{Derjaguin}).  Indeed, \eq{eq:Lw} can be written also as
\begin{equation}
  \mathcal{L}^{(W)} (\theta,\varphi) := \frac{1}{\beta \eta}
  \int_0^\infty dx\; x\left[ 
    1 - \mathrm{e}^{-\beta W (R+x,\theta,\varphi)} 
  \right] ,
\end{equation}
up to corrections of order $(\beta W)^2$, which are neglected in the 
approximation of small $W$ employed in our calculations.

\section{Details of the illustrative example in Figs.~1 and 2}
\label{sec:examples}
  
Here we collect the specific details for the example introduced in the
main text consisting of a particle with an activity which comprises mono-, 
di-, and quadrupolar contributions:
\begin{equation}
 \label{eq:act_particular}
 \mathbb{A}(\theta,\phi) = a_{00} Y_{0,0}(\theta,\phi) + a_{10} 
Y_{1,0}(\theta,\phi) + a_{20} Y_{2,0}(\theta,\phi)\, .
\end{equation}
(The values $a_{1,\pm 1}=0$ can be realized via a convenient choice of
the orientation of the $z$--axis, and thus it does not represent a
loss of generality.)  In this case, \eq{eq:driftV} renders (with
$\tilde{\xi} := \xi/R$)
\begin{eqnarray}
 \label{Vz_partic}
  \frac{V_z}{V_0}
  & =
  & a_{00} a_{10} \left[
    g_{01}^\perp(\tilde{\xi}) (\bG_{00;10}^\perp \cdot \be_z)
    + g_{01}^\parallel(\tilde{\xi}) (\bG_{00;10}^\parallel \cdot \be_z)
    + g_{10}^\perp(\tilde{\xi}) (\bG_{10;00}^\perp \cdot \be_z)
    + g_{10}^\parallel(\tilde{\xi}) (\bG_{10;00}^\parallel \cdot \be_z)
    \right] 
    \nonumber \\
  &+
  & a_{10} a_{20} \left[
    g_{12}^\perp(\tilde{\xi}) (\bG_{10;20}^\perp \cdot \be_z)+ 
    g_{12}^\parallel(\tilde{\xi}) (\bG_{10;20}^\parallel \cdot \be_z)+
    g_{21}^\perp(\tilde{\xi}) (\bG_{20;10}^\perp \cdot \be_z)+ 
    g_{21}^\parallel(\tilde{\xi}) (\bG_{20;10}^\parallel \cdot \be_z)
    \right]\nonumber\\
  &=
  & \frac{a_{00} a_{10}}{\sqrt{3}} \left[
    g_{01}^\perp(\tilde{\xi}) 
    - g_{01}^\parallel(\tilde{\xi}) 
    + g_{10}^\perp(\tilde{\xi})
    \right] 
       + \frac{a_{10} a_{20}}{\sqrt{15}} \left[
    2 g_{12}^\perp(\tilde{\xi}) 
    - 2 g_{12}^\parallel(\tilde{\xi}) 
    + 2 g_{21}^\perp(\tilde{\xi}) 
    + g_{21}^\parallel(\tilde{\xi}) 
    \right]\nonumber\\
&:= & a_{10} \left[a_{00} f_1(\tilde{\xi}) + a_{20} f_2(\tilde{\xi})\right]\,.
\end{eqnarray}
Using Eqs. (\ref{eq:Gr})-(\ref{eq:gp}) and (\ref{eq:psiell}), one
arrives at
\begin{equation}
  \label{eq:f1example}
  f_1(\tilde{\xi})= \frac{\mathrm{e}^{\tilde{\xi}^{-1}} \left(12 \tilde{\xi} ^2-1\right)\,
    \text{Ei}(-\tilde{\xi}^{-1})-\tilde{\xi}  \left(6 
\tilde{\xi}^3-10 \tilde{\xi}^2-\tilde{\xi} +1\right)}{32 \sqrt{3} \, \tilde{\xi}  
\, (\tilde{\xi}
+1) \, [2 \tilde{\xi}  (\tilde{\xi} +1)+1]},
\end{equation}
where Ei$(x)$ denotes the exponential integral function
\cite{abram,GrRy94}, and
\begin{widetext}
  \begin{align}
    \label{eq:f2example}
    f_2(\tilde{\xi}) =
    \resizebox{.9\hsize}{!}{$\frac{\tilde{\xi} \, \boldsymbol{\{}\tilde{\xi}
        \, \boldsymbol{[}\tilde{\xi} \, \boldsymbol{(}4 \tilde{\xi} \, \{3
        \tilde{\xi} \, [\tilde{\xi} \, (7 \tilde{\xi}-5)
        -11] -5\}-25\boldsymbol{)}
        +3\boldsymbol{]}+2\boldsymbol{\}}-\mathrm{e}^{\tilde{\xi}^{-1}}
        \, \boldsymbol{[} \tilde{\xi} \, \boldsymbol{(} 6 \tilde{\xi}
        \; \{ \tilde{\xi} \, [24 \tilde{\xi}
        (\tilde{\xi} +1)+7]+4\}-5 \boldsymbol{)} - 2 \boldsymbol{]}
        \, \text{Ei}(-\tilde{\xi}^{-1})}{192 \sqrt{15} \,
        \tilde{\xi} ^2 \, [2 \tilde{\xi}\, (\tilde{\xi} +1)+1] \, \{\tilde{\xi}
    \, [9 \tilde{\xi} \, (\tilde{\xi} +1)+4]+1\}}$}.
  \end{align}
\end{widetext}
These functions are used to obtain Figs.~\ref{fig:Vz} and
\ref{fig:reentr}, and feature in the related discussions in the main
text. These explicit results allow one to study the asymptotic limits
$\tilde{\xi}\to 0$ and $\tilde{\xi}\to\infty$, respectively, and to
compare with the general results obtained previously. By using the
asymptotic behavior of the exponential integral function in
Eqs.~(\ref{eq:f1example}) and (\ref{eq:f2example}), one obtains
$V\sim\tilde{\xi}^5$ when $\tilde{\xi}\ll 1$ (in agreement with the
conclusions from Sec.~\ref{sec:Lambda0}), and $V\sim\tilde{\xi}^0$
when $\tilde{\xi}\gg 1$ (in agreement with the conclusions from
Sec.~\ref{sec:Lambdainfty}).

%